\documentclass[aps,twocolumn,amsmath,amssymb,pre,showpacs]{revtex4}

\usepackage[dvips]{graphicx}

\begin{document}

\title{Fractal Properties of Anomalous Diffusion in Intermittent Maps}
\author{Nickolay Korabel$^{1}$, Rainer Klages$^{2}$, Aleksei V.\ Chechkin$^{3}$,  
Igor M.\ Sokolov$^{4}$, Vsevolod Yu.\ Gonchar$^{3}$}
\affiliation{$^1$Courant Institute of Mathematical Sciences, New York
University, 251 Mercer Street, New York, NY 10012, USA\\
$^2$School of Mathematical Sciences, Queen Mary, University of
London, Mile End Road, London E1 4NS, UK\\
$^3$Institute for Theoretical Physics NSC KIPT,
Akademicheskaya st.1, 61108 Kharkov, Ukraine\\
$^4$Institute for Physics, Humboldt-Universit\"at zu Berlin, Newtonstra{\ss}%
e 15, D-12489 Berlin, Germany}

\date{\today}

\begin{abstract}
  An intermittent nonlinear map generating subdiffusion is
  investigated. Computer simulations show that the generalized
  diffusion coefficient of this map has a fractal, discontinuous
  dependence on control parameters. An amended continuous time random
  walk theory well approximates the coarse behavior of this quantity
  in terms of a continuous function. This theory also reproduces a
  full suppression of the strength of diffusion, which occurs at the
  dynamical transition from normal to anomalous diffusion.
  Similarly, the probability density function of this map exhibits a
  nontrivial fine structure while its coarse functional form is
  governed by a time fractional diffusion equation. A more detailed
  understanding of the irregular structure of the generalized
  diffusion coefficient is provided by an anomalous Taylor-Green-Kubo
  formula establishing a relation to de Rham-type fractal functions.
\end{abstract}

\pacs{05.45.Ac, 05.60.-k, 05.40.Fb}

\maketitle

\section{Introduction}

Since more than a century normal diffusion, characterized by a linear
increase in time of the mean square displacement (MSD) of an ensemble
of moving particles, provides a paradigmatic example of a stochastic
process. Here the MSD can be expressed by the ensemble average
$\left<x^2\right> \sim t^\beta$ with an exponent $\beta=1$, where $x$
holds for the position of a particle on the real line at time $t\ge0$.
However, exponents $\beta\neq1$ are also possible yielding two
important examples of anomalous diffusion: superdiffusion with $\beta
> 1$ and subdiffusion with $0<\beta < 1$. More recently, the
importance of anomalous diffusive regimes was realized not only in
physics but also in chemistry, biology and economics. These regimes
were found in theoretical models and experiments related, among
others, to turbulence, amorphous semiconductors, porous media, surface
diffusion, glasses, granular matter, reaction-diffusion processes,
plasmas and biological cell motility, see
Refs.~\cite{Bou90,Shl93,Shl94,Klaf96,Met00,Sok02,Zas02,Met04,EbSo05}
for reviews. Parallel to this development low dimensional
deterministic maps attracted much attention as simple models of
anomalous dynamics which can be understood analytically
\cite{Pom80,Gei84,Gei85,Shl85,Gas88,Wang,Art93,Zum93,Zum93a,Zum95,Sto95,Det97,Dra00,Gas02,Bar03,Art03,Rad04}.
A characteristic feature of the stochastic processes generated by
these maps is provided by the probability density functions (PDFs) of
the dynamical variables. In case of normal diffusion these PDFs
exhibit Gaussian forms, whereas for subdiffusion they yield tails with
stretched exponential decay and for superdiffusion L{\'e}vy power laws
\cite{Zum93}.

Several theoretical approaches have been worked out over the past few
decades in order to explain anomalous diffusion. Perhaps the most
famous one is continuous time random walk (CTRW) theory, pioneered by
the work of Montroll, Weiss and Scher \cite{MW}. Their stochastic
approach was later on adapted to sub- and superdiffusive deterministic
maps \cite{Gei84,Gei85,Shl85,Zum93,Zum93a,Zum95,Dra00,Bar03}. Related
maps were originally proposed by Pomeau and Manneville as simple
models of intermittency \cite{Pom80}. A classification of their
dynamics was provided particularly by Gaspard and Wang combining
stochastic with dynamical systems theory \cite{Gas88}.  Exploiting the
deterministic properties of these maps, anomalous diffusion was
further studied in the framework of the thermodynamic formalism
\cite{Wang,Sto95}, by means of periodic orbit theory
\cite{Art93,Det97,Art03} and by spectral decomposition techniques
\cite{Gas02}. Currently aging phenomena \cite{Bar03,Rad04},
non-ergodic behavior \cite{Lutz04} and infinite invariant measures
\cite{Hu95} are in the focus of investigations demonstrating that
these simple models provide ongoing inspiration for important new
research.

However, all of the above studies focused on specific values of
control parameters only for which these maps are to a large extent
analytically tractable. That generally the dynamics is more intricate
was shown by calculating the parameter-dependent diffusion coefficient
of a piecewise linear one-dimensional map, which turned out to be
fractal \cite{Kla95,Kla96,GrKl02}. Similar behavior was detected in
more complicated models like the climbing sine map \cite{Kor02}, in
experimentally accessible systems like particles bouncing on
corrugated surfaces \cite{Har01} and in models of Josephson junctions
\cite{Tan02}. These findings can be explained by the existence of
deterministic dynamical correlations that are topologically unstable
under parameter variation \cite{Kla95,Kla96,KlKo02}; for a review see
Part 1 of Ref.\ \cite{Kla04}.

In this paper we show that a fractal parameter dependence of physical
quantities is not only typical for low-dimensional periodic
deterministic dynamical systems exhibiting normal transport laws but
also for anomalous dynamics. As an example we study
parameter-dependent subdiffusion in a simple deterministic map,
however, our main arguments hold for a broad class of anomalously
diffusive systems. A specific feature of our analysis is that it
employs a blending of techniques from stochastic theory and the theory
of dynamical systems.  That way we wish to contribute to the
microscopic foundations of a general theory of anomalous deterministic
transport, which does not yet seem to be fully developed.

Our paper is organized as follows: The model is introduced in Section
\ref{model}. As a rate of diffusivity we choose a generalized
diffusion coefficient (GDC), which generalizes the diffusion constant
known from normal diffusion.  In Section \ref{NumRes} we present
results from computer simulations showing that the GDC has a
nonmonotonic, irregular dependence on control parameters of the map.
We give a qualitative explanation of this phenomenon by arguing that
the GDC is a self similar-like, fractal function. The GDC is
furthermore conjectured to be everywhere discontinuous. In Section
\ref{theory} we briefly review CTRW theory. In Section
\ref{mod_theory} we show that an amended version of it correctly
describes the coarse functional form of the GDC. Apart from fractal
parameter dependencies, the map under consideration exhibits an
interesting dynamical transition, which is studied in detail in
Section \ref{sup}. In Section \ref{FFP} we compute the PDF of this map
from simulations and analyze it by means of a time fractional
diffusion equation. Section \ref{ff} finally outlines a deterministic
approach for analyzing the fractal GDC, which is based on a
Taylor-Green-Kubo (TGK) formula for anomalous diffusion.  This theory
enables us to relate anomalous diffusion processes to fractal de
Rham-type functions. Section \ref{con} summarizes our results. For an
outline of this work we refer to Ref.\ \cite{Unp}.

\section{The Model}\label{model}

\begin{figure}[t]
\centering
\includegraphics[width=0.4\textwidth]{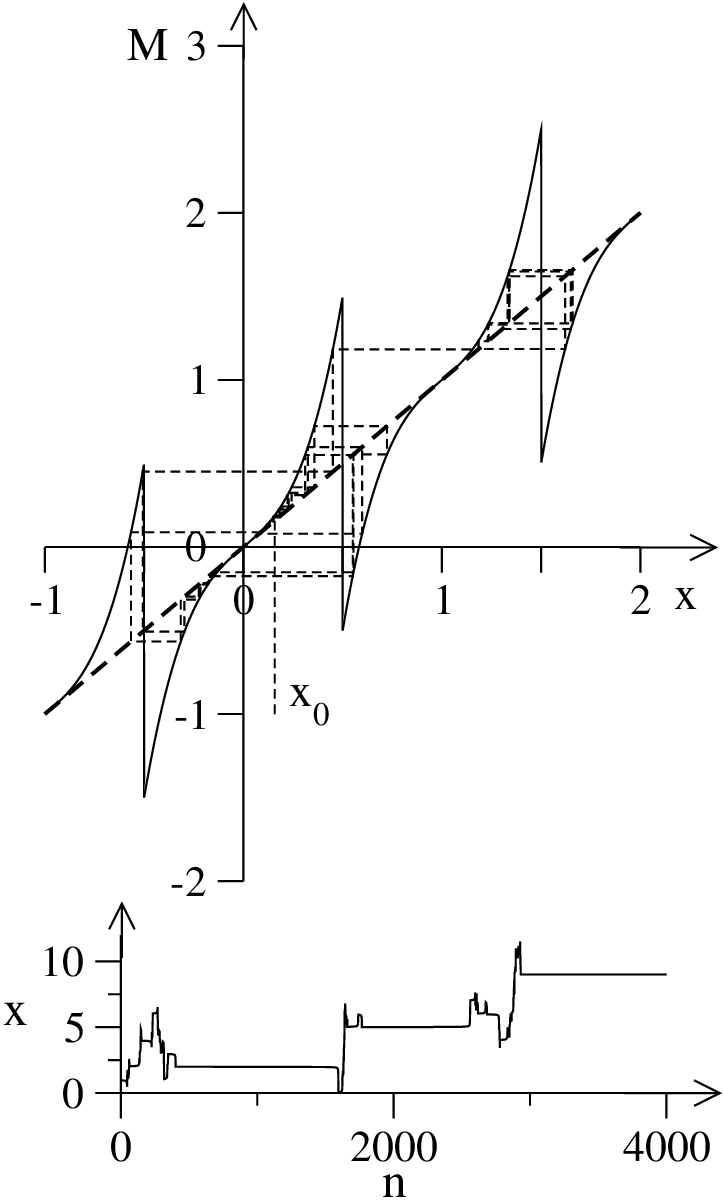}\\
\caption{Top: Illustration of the spatially continued map $M$ 
  Eq.~(\ref{map_eq}) for parameters $z=3, a=8$ with a typical
  trajectory generated by the map (cobweb plot with dashed horizontal
  and vertical lines) starting at initial condition $x_0$. Bottom: The
  demonstration of intermittency in a time series plot of a typical
  trajectory. Note the long laminar periods in the trajectory, which
  are interrupted by chaotic `bursts'. Here and in the following
  figures all quantities are without units.}
\label{FigMap}
\end{figure}

We focus on a subdiffusive map which, restricted onto the unit
interval, was introduced by Pomeau and Manneville to describe
intermittency \cite{Pom80},
\begin{equation}
x_{n+1} \equiv M_{z,a}(x_n) = x_n + a x_n^z, \; \; 0 \le x_n < \frac{1}{2}\quad ,
\label{map_eq}
\end{equation}
where the parameter $z\ge 1$ holds for the degree of nonlinearity and
$a\ge 1$ is a second control parameter. In the following we will omit
these two indices for convenience, $M_{z,a} \equiv M$. The variable
$x_n\in\mathbb{R}$ denotes the position of a point particle at
discrete time $n\in\mathbb{N}_0$. Translation symmetry, $M(x + 1) =
M(x) + 1$, and reflection symmetry, $M(-x) = - M(x)$, complete the
definition of the map on the real line.  The iterated dynamics of this
model is illustrated in Fig.\ \ref{FigMap} for particular values of
the two control parameters: Given some initial condition $x_0$, the
equations of motion Eq.~(\ref{map_eq}) produce the next value $x_1$,
$x_1$ determines $x_2$, and so on. A typical trajectory of the system,
displayed in Fig.\ \ref{FigMap} in a cobweb plot, is therefore
represented in form of jumps $x_0\rightarrow x_1 \rightarrow x_2
\rightarrow \ldots$ on the real line.
\begin{figure}
\centering
\includegraphics[width=0.48\textwidth]{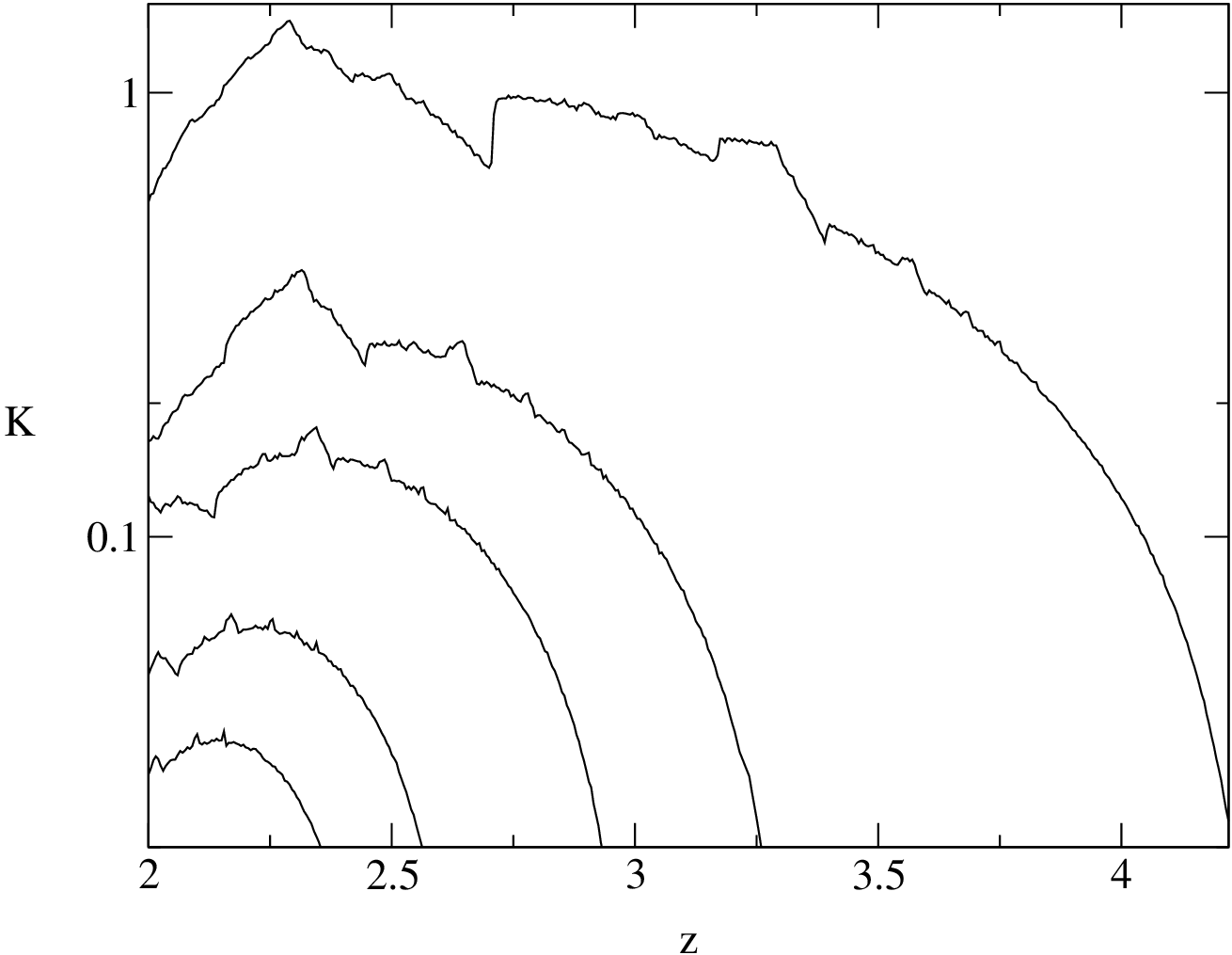}\\
\caption{The generalized diffusion coefficient 
  $K$, Eq.~(\ref{K}), for the map Eq.~(\ref{map_eq}) as a function of
  the map's nonlinearity $z$ for different values of the prefactor $a$
  in the anomalous diffusive region $z>2$. The semi-logarithmic scale
  is chosen in order to magnify the irregular fine structure.  The
  values for $a$ from bottom to top are $a=2.78, 3.14, 4, 5, 9.8$. The curves consist of $100, 130, 200, 254, 450$ points, respectively.}%
\label{K_z_A}
\end{figure}

For $1 \le z < 2$ the diffusion process generated by this map is
normal, whereas for $z \ge 2$ it is anomalous
\cite{Gei84,Gas88,Zum93}.  This anomaly is due to the existence of
marginally stable fixed points located at all integer values of $x_n$
around which a particle gets `trapped' for a long time \cite{Wang}.
However, in regions with larger local derivative, where the particle
can `jump` to different unit intervals, the dynamics becomes more
irregular, see Fig.\ \ref{FigMap}. Thus, a typical trajectory of the
spatially continued map Eq.~(\ref{map_eq}) consists of long laminar
pieces interrupted by short chaotic bursts. Such a behavior is the
hallmark of what is called intermittency \cite{Pom80}. An example of a
typical intermittent trajectory generated by this map is shown at the
bottom of Fig.\ \ref{FigMap}.

In the groundbreaking works of Thomae, Geisel \cite{Gei84}, Zumofen
and Klafter \cite{Zum93}, main attention was paid to the time
dependence of the MSD for specific fixed values of the parameter
$a$. In contrast, our aim is to study how a suitably defined GDC
behaves under variation of the two control parameters $a$ and $z$. In
Refs.\ \cite{Met00,KDM05} an anomalous diffusion coefficient was
introduced by 
\begin{equation} 
\label{GDC_Def} 
D\equiv D(z,a) = \lim_{n \rightarrow \infty} \frac{\Gamma (1 + \beta)}{2} \frac{\left<
x^2 \right>}{n^{\beta}} \quad , 
\end{equation} 
where $\Gamma$ is the gamma function and $\beta$ is some constant that will be specified in
the following. For convenience we define the GDC by 
\begin{equation}
\label{K} 
K\equiv K(z,a) = \frac{2 D}{\Gamma (1 + \beta)} \quad .
\end{equation} 
For numerical simulations of the spatially extended map
Eq.~(\ref{map_eq}) we have typically used an ensemble of $10^6$
particles, where the initial conditions are uniformly distributed on
the unit interval $(0,1)$. If not said otherwise, each trajectory was
calculated up to $10^4$ time steps. For different values of $a$ we
have confirmed that the numerically computed power law dependence of
the MSD is in full agreement with the CTRW solution \cite{Zum93}
\begin{equation}
\label{alfa} \beta = \begin{cases} 1 , & 1 \le z < 2 \cr
\frac{1}{z-1}, & z \ge 2\quad .  \end{cases} 
\end{equation} 
We remark that at the transition point between normal and anomalous
diffusion a logarithmically corrected dependence of the MSD is
obtained,
\begin{equation} 
\left<x^2\right>\sim n/\ln n \quad \mbox{for}\quad
z=2 \quad , \label{supp} 
\end{equation} 
which again is in complete agreement with CTRW theory. Hence, for the
rest of this paper $\beta$ is defined by the theoretically correctly
predicted values Eq.~(\ref{alfa}).

\section{Numerical results}\label{NumRes}

\begin{figure}[t] 
\centering
  \includegraphics[width=0.48\textwidth]{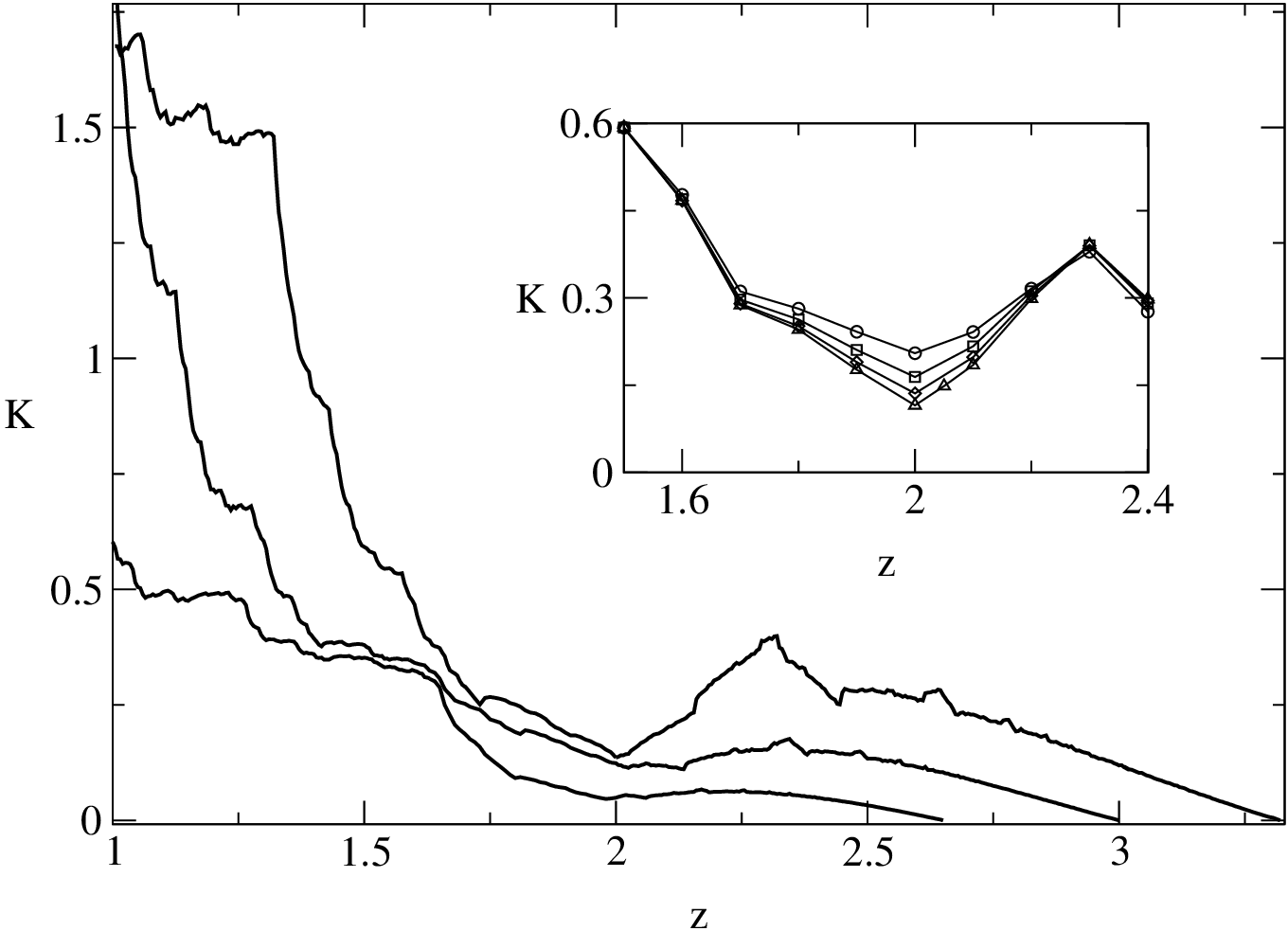}\\ 
  \caption{The generalized diffusion coefficient $K$ as a function of
    $z$ for different values of $a$. The parameter values from bottom
    to top (at $z=2$) are $a=3.14, 4, 5$. The curves consist of
      $296,360,435$ points, respectively.  The inset shows several
      representative values of $K$ around $z=2$ for $a=5$. Lines are
      guides for the eyes only.  Different symbols from top to bottom
      correspond to simulations with $10^6$ trajectories and the
      different computation times $n=10^2$ (circles), $10^3$
      (squares), $10^4$ (rhombuses), $10^5$ (triangles) indicating
      ultra-slow convergence of $K$ towards zero at $z=2$. Also the
      curve calculated with $10^7$ trajectories, each of the length of
      $10^4$ time steps, is shown. In the inset it coincides with the
      curve calculated with $10^6$ trajectories, each of the length of
      $10^4$ time steps.}
\label{K_z}
\end{figure}

We first fix the parameter $a$ and study the dependence of $K$ as a
function of the degree of nonlinearity $z$. Numerical results are
presented in Figs.\ \ref{K_z_A} and \ref{K_z}. $K$ as a function of
$z$ appears to be highly irregular exhibiting a lot of structure on
fine scales. Particularly Fig.~\ref{K_z_A} shows how the fine
structure as a function of $z$ evolves under variation of $a$.

In Fig.\ \ref{K_z} the calculations were performed both for the normal
and for the anomalous diffusive regime. Eq.~(\ref{supp}) suggests that
at $z=2$ the GDC exhibits a local minimum in form of a total
suppression of the strength of anomalous diffusion, that is,
$K(2,a)=0$ for all values of $a$. Interestingly, this transition from
normal to anomalous diffusion appears to be approached continuously by
$K$ as a function of $z$. The reason is that for $z\neq 2$ logarithmic
terms are still present in the dynamics, but they contribute for
transient times only. This peculiar behavior does not only diminish
the GDC but also significantly slows down the convergence of the
simulation results. The inset of Fig.\ \ref{K_z} shows several
  representative values of $K$ in the transition region, where
  different symbols correspond to different computation times and
  different numbers of trajectories for fixed $a$. Due to the slow
convergence, even for the largest computation times the results for
$K$ are quantitatively still apart from the CTRW prediction, which
holds in the limit of time to infinity. However, qualitatively there
is a tendency towards zero at $z=2$. This specific problem will be
analyzed in full detail in Section \ref{mod_theory}.

\begin{figure}[t]
\centering
\includegraphics[width=0.5\textwidth]{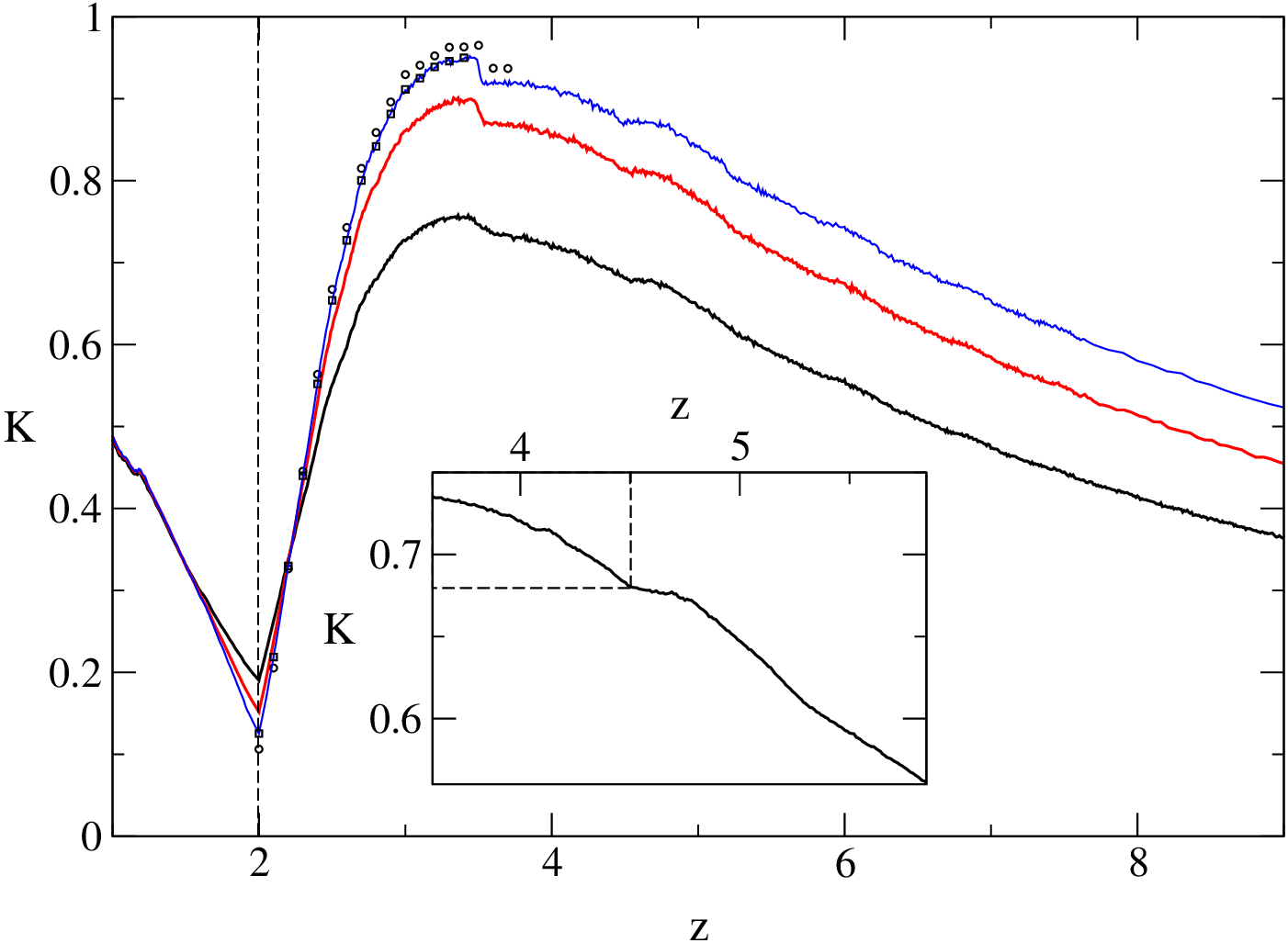}\\
\caption{The generalized diffusion coefficient $K$ as a function of 
$z$ for the fixed height of the map $h=\sqrt{3}$ as defined in the text. 
Different curves correspond to simulations with $10^6$ 
trajectories and different computation times, from bottom to top 
on the right hand side of the figure $n=10^2, 10^3, 10^4$. Note 
the inverse order of the curves at the point $z=2$ marked by the 
vertical dashed line. Curves consist of $854, 646, 650$ points, 
respectively. Several values of $K$ for $n=10^5$ are also shown (circles). 
The data indicate clear convergence in time and the 
formation of irregularities. Additionally, several representative 
values of $K$ calculated for $10^7$ trajectories, each of the 
length of $10^4$ time steps, are shown (squares). They coincide 
with corresponding values of $K$ calculated with $10^6$ 
trajectories, each of the length of $10^4$ time steps. The inset 
shows a magnified region of the curve calculated for $n=10^2$. The 
dashed box marks a region that appears to be self-similar to the 
whole region shown in the inset indicating the existence of an 
underlying fractal pattern.  The curve was calculated for $2 \cdot 10^7$ 
trajectories and consists of $200$ points.}
\label{K_h}
\end{figure}

From normal diffusive maps it is known that iterations of the critical
points of a map, which here are the points of discontinuity at
$x=1/2+m\:,\:m\in\mathbb{Z}$, play a crucial role in order to
understand the complicated parameter dependence of a diffusion process
\cite{Kla95,Kla96}. Accordingly, variations of the height of the map
$h:= M(\frac{1}{2}) = \frac{1}{2} + a (\frac{1}{2})^z$ strongly affect
the parameter-dependent GDC, as we will discuss in detail below. Such
variations are achieved both by changing $z$ and $a$. In order to more
clearly represent the impact of variations of $z$ only on the GDC, we
study $K$ as a function of $z$ for fixed height $h$. We 
emphasize that the topology of the orbit of the critical point is 
still affected by variations of $z$ only, however, it is less 
sensitive to $z$ than to varying $h$. Respective simulation results
are shown in Fig.\ \ref{K_h}. As expected, $K(z,h)$ for fixed $h$ is
considerably smoother compared to Figs.\ \ref{K_z_A} and \ref{K_z}.
Indeed, the local minimum in the transition from normal to anomalous
diffusion is now less obscured by a non-trivial fine structure.
However, note that very small irregularities forming a self 
similar-like pattern are still present, as is suggested by 
Fig.\ \ref{K_h} and the inset, which depicts a magnification of a small 
characteristic region. The figure also shows the convergence of the 
data with respect to simulation time and number of trajectories 
from which the formation of the irregularities is clearly seen.

\begin{figure}[t]
\centering
\includegraphics[width=0.48\textwidth]{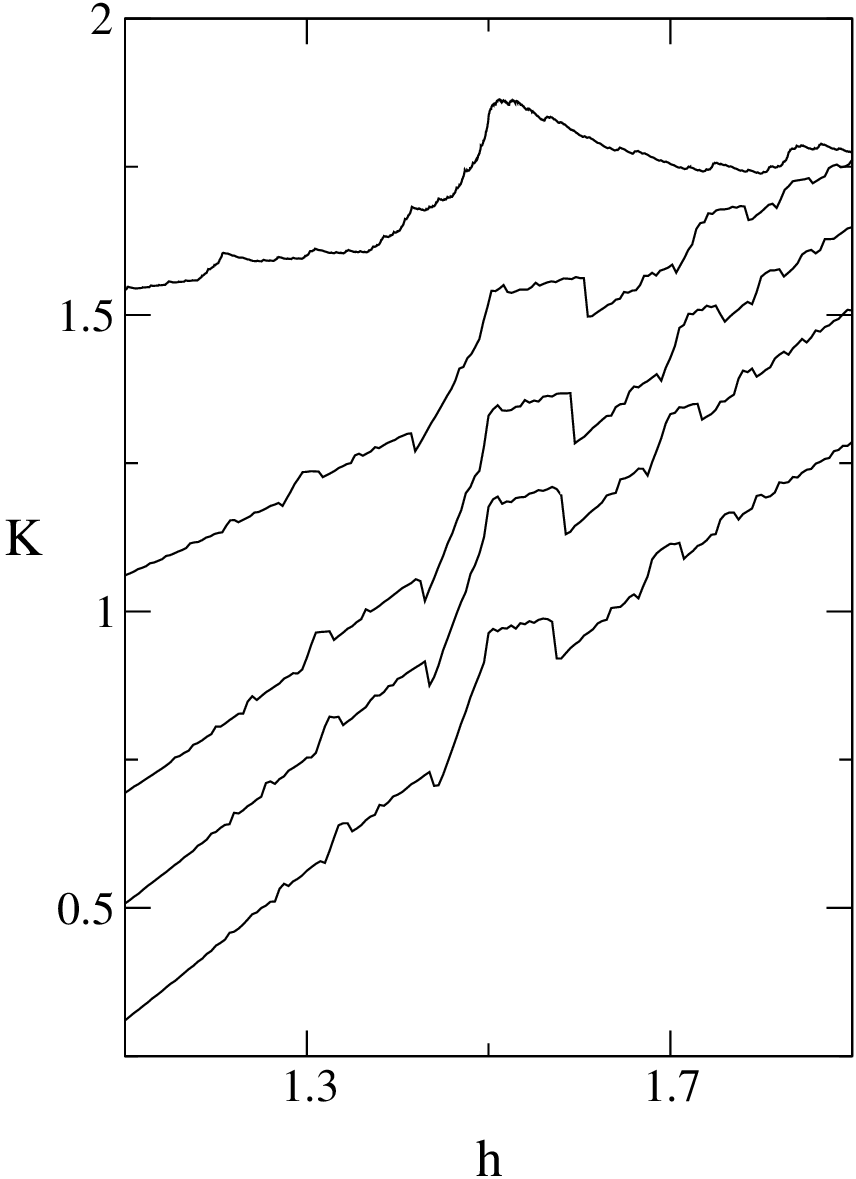}\\ 
\caption{The generalized diffusion coefficient $K$ as a function of
parameter $h$ for different values of $z$. From top 
to bottom it is $z=1, 2.5, 3, 3.5, 4$. For clarity the curves are shifted vertically by
$1.5, 1, 0.6, 0.4, 0.2$, respectively.}
\label{K_s} 
\end{figure}

\begin{figure}[t] \centering
\includegraphics[width=0.49\textwidth]{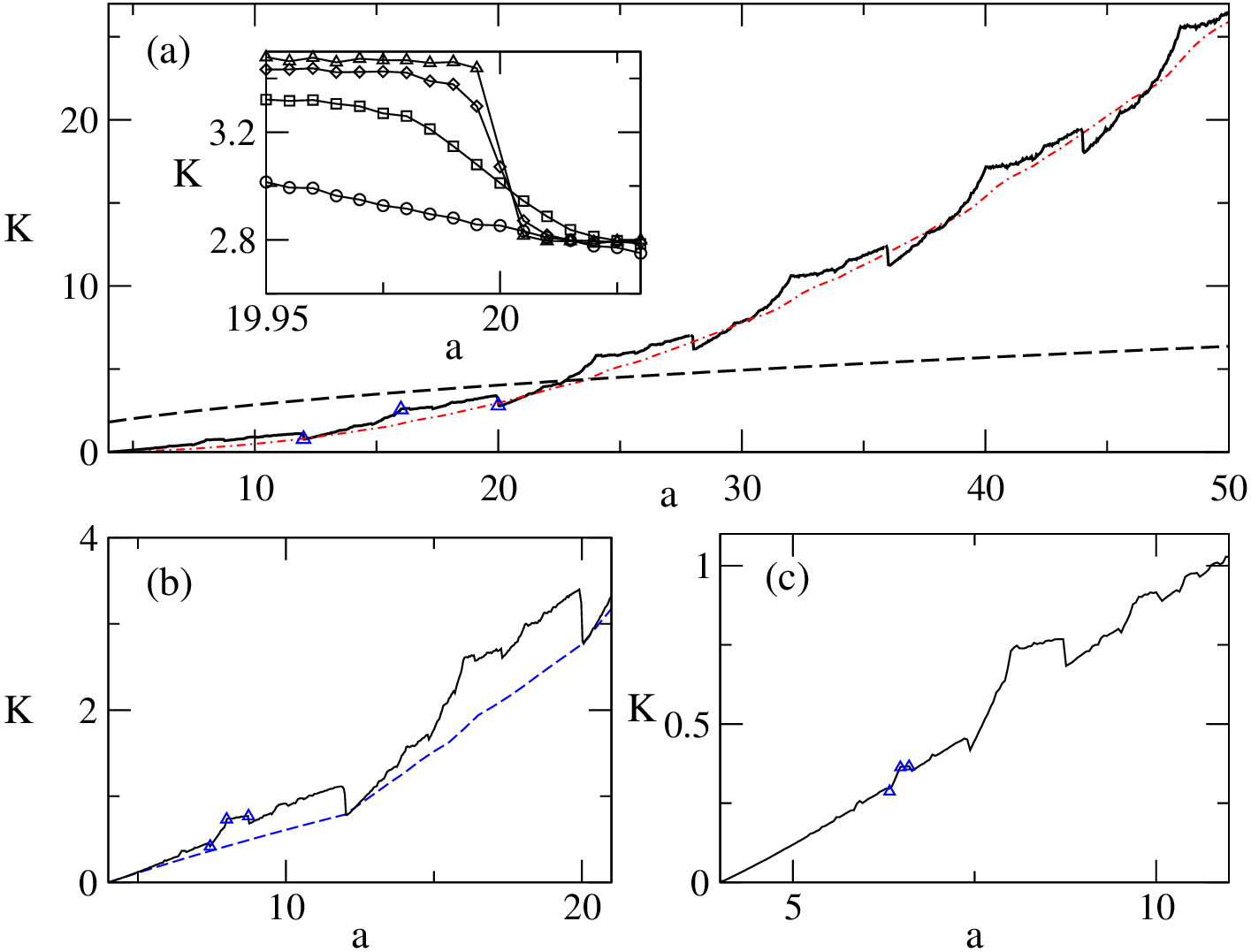}\\
\caption{(a) The generalized diffusion coefficient
  $K$ over a large range of the parameter $a$ at $z=3$. Note that
  $K(3,a)=0$ for $1\le a\le 4$. The dashed line represents the CTRW
  approximation Eq.~(\ref{CTRW_Geisel}). The dashed-dotted line is the
  modified CTRW Eqs.~(\ref{jump_length}), (\ref{mod_GDC_CTRW}). (b)
  and (c) show magnifications of the initial region of (a)
  demonstrating the existence of irregularities on finer scales. The
  dashed line in (b) shows the modified CTRW Eqs.
  (\ref{int_jump_length}), (\ref{mod_GDC_CTRW}). The sets of symbols
  included in (a) to (c) correspond to specific turnstile parameter
  values (see Section \ref{NumRes} for explanations). The inset in (a)
  depicts the development of a discontinuity in the GDC around $a=20$.
  The different lines correspond to different iteration times, 
  from bottom to top at left $n=10^3$ (circles), $10^4$ (squares), 
  $10^5$ (rhombuses), $10^6$ (triangles), thus yielding a higher 
  numerical precision. The curve with $10^7$ trajectories and $10^4$ 
  time steps, which coincides with the one calculated with $10^6$ 
  trajectories and $10^4$ time steps, shows that we gain convergence 
  with respect to the number of trajectories. The data in (a) 
  consists of $1198$ points.}
\label{Z3}
\end{figure}

Now we focus on the dependence of $K$ on the parameter $h$ with
the parameter of nonlinearity $z$ kept fixed. In Fig.\ \ref{K_s} the
behavior of the GDC as a function of $h$ is shown for different values
of $z$ which, as expected, is more non-monotonous than that of $K(z)$. 
The structure of $K(h)$ for different $z$ looks somewhat
similar, however, we could not detect a simple scaling. For $z=1$ the
map Eq.~(\ref{map_eq}) reduces to a piecewise linear map for which it
was shown analytically and numerically that diffusion is normal and
that the diffusion coefficient is a fractal function of the slope $a$
\cite{Kla95,Kla96,Klauss,GrKl02,KlKo02,Kla04}, see the upper curve in
Fig.\ \ref{K_s}. For $z=3$ and over a larger range of $a$, $K(a)$ as
obtained from simulations is presented in Fig.\ \ref{Z3} (a).
Magnifications of the initial region, see Fig.\ \ref{Z3} (b) and (c),
show a self similar-like behavior of $K$ on finer and finer scales
indicating a fractal structure.

A qualitative explanation for the fractality of all these curves can
be given in terms of {\it turnstile dynamics}, which has successfully
been applied in order to understand the fractality of the normal
diffusion coefficient in piecewise linear maps. Here we outline the
basic idea of this approach only, for technical details we refer to
Refs.\ \cite{Kla95,Kla96}. The key ingredient of this method is the
orbit generated by iterations of the critical point $x_c=1/2$ of the
map $M$ restricted onto the unit interval, $x_n=M^n(x_c)
\:\mbox{mod}\: 1$.  If this orbit is periodic it defines a {\it Markov
partition} with certain partition parts representing coupling regions
({\it turnstiles}) where, in the associated periodically continued
map, particles can jump from one unit interval to another
one. Specific parameter values define Markov partitions with specific
turnstile couplings which, in more physical terms, generate specific
sequences of forward- and backward scattering of particles that start
nearby the critical point.  It turns out that these Markov partitions
are topologically highly unstable under parameter variation, in the
present case both for varying $z$ and $a$. 

In Fig.\ \ref{Z3} (a) three parameter values are identified
representing Markov partitions that correspond to two local minima and
a local maximum of $K(a)$ thus highlighting a specific fine structure
in the whole curve. Higher-order Markov partitions can now be
constructed (generated by higher-order iterations of the critical
point) which are of the same type as the initial three identified in
Fig.\ \ref{Z3} (a), however, yielding new sets of parameter values.
Such higher-order parameter triples are shown in Fig.\ \ref{Z3} (b),
(c). They identify the same type of fine structure on finer and finer
scales. A similar analysis can be performed for other local peaks of
$K(a)$ qualitatively explaining the fractal structure of the GDC.
More quantitatively, one may wish to calculate some fractal dimension
for these curves.  However, this is a hard task even if analytical
formulas for a diffusion coefficient are available \cite{Klauss} and
is left for further studies.

We conclude this section with an unexpected detail of the GDC. For the
piecewise linear map at $z=1$ the diffusion coefficient as a function
of slope of the map is continuous \cite{GrKl02}. For $z=3$, however,
numerical results suggest that the dependence of $K$ on $a$ is
discontinuous. In the inset of Fig.\ \ref{Z3} we have magnified a
small region of $K$ around $a=20$ (analogous behavior is found around
$a=12,28,36,\ldots$). Note that for $a>20$ particles can jump for the
first time from the first unit interval to the fourth one. Increasing
the numerical precision by increasing the computation time, $K(a)$
near $a=20$ gradually approaches a step function, which suggests that
a discontinuity develops. Similar discontinuities have already been
reported for Lyapunov exponents as functions of a bias in piecewise
linear maps \cite{KlCP}. More complicated ones are known to exist for
diffusion coefficients of nonlinear maps exhibiting bifurcation
scenarios \cite{Kor02} and have recently been highlighted in research
on transport in polygonal billiard channels \cite{JeRo06}. An
  explanation why for our model the GDC is discontinuous for $z=3$
  while it is continuous for $z=1$ may be sought in the completely
  different character of the PDF of the map for $z=3$, which develops
  non-integrable singularities at all marginally stable fixed points.
  Details of this peculiar behavior will be discussed later, see
  Fig.~\ref{PDF} and the analysis after Eq.~(\ref{MSD2}). Note that
  at suspected points of discontinuity of the GDC like $a=20$ the
  critical point is getting mapped right onto these singular regions
  of the PDF at the very first iteration. Simultaneously, the orbit of
  the critical point exhibits a transition from forward to backward
  scattering under variation of $a$, which leads to a local maximum of
  the GDC in terms of turnstile dynamics. For $z=3$ this change in
  the microscopic scattering process is drastically amplified by the
  singular behavior of the PDF probably leading to a discontinuity in
  the GDC. In contrast, for $z=1$ the PDF is not singular but a
  non-differentiable step function \cite{Kla96}, thus the GDC is
  locally maximal but still continuous. It would be valuable to have a
  mathematical proof of this conjectured discontinuity phenomenon.

The discontinuity at $a=20$ is furthermore a boundary point of the
specific fine structure identified in Fig.~\ref{Z3}, which was
previously discussed in terms of turnstile dynamics.  As we have just
emphasized, this parameter value is determined by a specific periodic
orbit of the critical point of the map. However, as was outlined
before, one can now construct infinitely many higher-order iterates of
the same type of periodic orbits yielding new sets of parameter
values. These parameter values are related to the same type of
structure and, hence, identify the same type of discontinuity on finer
and finer scales.  Furthermore, such parameter values are typically
densely distributed on the parameter axis \cite{Kla95,Kla96}. This
leads us to conjecture that $K(a)$ shown in Fig.~\ref{Z3} is 
discontinuous on a dense set of parameter values (probably being of
  measure zero). Note that the argument carries over to $K(z)$
  in Fig.~\ref{K_h}, where a careful study of the largest point of
  irregularity at $z\simeq 3.4$ reveals the same type of
  discontinuity.

So far we have focused on the fine structure of the GDC only. The next
section proceeds with an understanding of its coarse functional form.
As far as we know there is only one analytical result in the
literature trying to predict the whole parameter dependence of the GDC
of this map \cite{Gei84}. However, as we will show the respective
calculation needs to be modified in order to match to computer
simulation results. This necessitates to briefly review the whole
approach by explaining our corrections.

\section{Continuous time random walk theory for maps}\label{theory}

The CTRW theory of Montroll, Weiss and Scher \cite{MW} has become a
standard tool to model diffusion in intermittent maps like
Eq.~(\ref{map_eq}) \cite{Gei84,Zum93}. This approach assumes that
diffusion can be decomposed into two stochastic processes
characterized by waiting times and jumps. Thus one has two sequences
of independent identically distributed random variables, namely a
sequence of positive random waiting times $T_1, T_2, T_3, \ldots$ with
PDF $w(t)$, $\int_{0}^{\infty} w(t) dt = 1$, and a sequence of random
jumps $\zeta_1, \zeta_2, \zeta_3, \ldots$ with a PDF $\lambda(x)$,
$\int_{-\infty}^{\infty} \lambda(x) dx = 1$. For example, if a
particle starts at point $x = 0$ at time $t_0 = 0$ and makes a jump of
length $\zeta_n$ at time $t_n = T_1 + T_2 + ... + T_n$, its position
is $x = 0$ for $0 \le t < T_1 = t_1$ and $x = \zeta_1 + \zeta_2 + ...
+ \zeta_n$ for $t_n \le t < t_{n+1}$. The probability that at least
one jump is performed within the time interval $[0,t)$ is then
$\int_0^t dt' w(t')$ while the probability for no jump during this
time interval reads $\Psi(t) = 1 - \int_{0}^{t} dt' w(t')$. The master
equation for the PDF $P(x,t)$ to find a particle at position $x$ and
time $t$ is then \begin{eqnarray}
  \label{master_eq} P(x,t) &=& \int_{-\infty}^{\infty} dx' \lambda (x
  - x') \int_{0}^{t} dt' \; w(t -
  t') \; P(x',t') + \nonumber\\
  & & + \Psi(t) \delta(x)\quad .
\end{eqnarray} 
It has the following probabilistic meaning: The PDF to find a particle
at position $x$ at time $t$ is equal to the PDF to find it at point
$x'$ at some previous time $t'$ multiplied with the transition
probability to get from $(x',t')$ to $(x,t)$ integrated over all
possible values of $x'$ and $t'$. The second term accounts for the
probability to remain at the initial position $x=0$. It is easy to
check that this equation yields a normalized PDF $P(x,t)$. The most
convenient representation of this equation is obtained in terms of the
Fourier-Laplace transform of the PDF,
\begin{equation} \label{times_PDF} 
\hat{\tilde{P}} (k,s) =
\int_{-\infty}^{\infty} dx \; e^{i k x} \int_{0}^{\infty} dt \;
e^{-st} P(x,t) \quad , 
\end{equation}
where the hat stands for the Fourier transform and the tilde for the
Laplace transform. Respectively, this function obeys the
Fourier-Laplace transform of Eq.~(\ref{master_eq}), which is called
the Montroll-Weiss equation \cite{MW},
\begin{equation} \label{Montroll_Weiss} 
\hat{\tilde{P}} (k,s) =
\frac{1 - \tilde{w}(s)}{s} \frac{1}{1-
\hat{\lambda}(k)\tilde{w}(s)}\quad .  \end{equation}
The Laplace transform of the MSD can be
readily obtained by differentiating the Fourier-Laplace transform
of the PDF,
\begin{equation}
\label{Laplace_MSD}
\tilde{ \left< x^2 (s) \right> } = \int_{-\infty}^{\infty} dx \; x^2 \tilde{P}(x,s) =
\left. - \frac{\partial^2 \hat{\tilde{P}} (k,s) }{\partial k^2} \right|_{k=0}\quad .
\end{equation}
In order to calculate the MSD within this theory, it thus suffices to
know $\lambda(x)$ and $w(t)$ generating the stochastic process. For
one-dimensional maps of the type of Eq.~(\ref{map_eq}), using the
symmetry of the map the waiting time distribution can be calculated
from the approximation
\begin{equation}
\label{cont_time_map}
x_{n+1} - x_n \simeq \frac{dx_t}{dt} = a x_t^z, \; \; x_t \ll 1 \quad ,
\end{equation}
where we have introduced the continuous time $t\ge 0$. Solving this
equation for initial condition $x_0$ yields
\begin{equation}
\label{cont_time_map_sol}
x_t = \left[\frac{1}{x_{0}^{z-1}} - a(z-1) t\right]^{-\frac{1}{z-1}}\quad .
\end{equation}
We now define that a particle makes a jump when it leaves the unit
interval. Note that this definition is different from the one of Ref.\ 
\cite{Gei84}, which used the interval $[-\frac{1}{2},\frac{1}{2}]$.
That the unit interval is the right choice for calculating the
parameter-dependent random walk diffusion coefficient was shown for
$z=1$ in Refs.\ \cite{Kla95,Kla96,KlDo97,KlKo02,Kla02}. From
Eq.~(\ref{cont_time_map_sol}) one can then obtain the time $t$ that a
particle spends on the unit interval before making a jump according to
the condition $x_t = 1$.  The waiting time thus becomes a function of
the injection point $x_0$,
\begin{equation}
\label{cont_time_map_sol3}
t(x_0) = \frac{1}{a(z-1)}\left(\frac{1}{x_{0}^{z-1}} - 1\right)\quad .
\end{equation}
Accordingly, the waiting time PDF $w(t)$ is related to the yet unknown
PDF of injection points,
\begin{equation}
\label{cont_time_map_sol4}
w(t) \simeq P_{in}(x_0) \left| \frac{dx_0}{dt}\right| \quad .
\end{equation}
Making the assumption that the PDF of injection points is uniform,
$P_{in} \simeq 1$, the waiting time PDF is calculated to
\begin{equation}
\label{times_PDF2}
w (t) = a \left[ 1 + a (z - 1) t \right]^{- \frac{z}{z-1}}
\end{equation}
for long enough times $t$, where normalization is used to obtain the
prefactor.

If jumps between neighboring cells {\em only} are taken into account, the
jump PDF and its Fourier transform may be assumed as \cite{Zum93}
\begin{equation}
\label{jump_PDF}
\lambda (x) = \delta (|x| - 1), \; \; \; \hat{\lambda}(k) = \cos (k)\quad .
\end{equation}
Combining Eq.~(\ref{Laplace_MSD}) with Eq.~(\ref{jump_PDF}) leads to
the Laplace transform \cite{Gei84}
\begin{equation}
\label{Laplace_MSD2}
\tilde{ \left< x^2 \right> } = \frac{\tilde{w}(s)}{s(1-\tilde{w}(s))} \quad .
\end{equation}
For the following calculations it is useful to define
\begin{equation}
\gamma:=\frac{1}{z-1}\:,\:z >1\quad . \label{eq:defgamma}
\end{equation}
Note that for $z\ge 2$, $\gamma$ is identical with $\beta$ defined in
Eq.~(\ref{alfa}).  For $\gamma\neq2$ the Laplace transform of Eq.\ 
(\ref{times_PDF2}) reads
\begin{equation}
\label{Laplace_times_PDF}
\tilde{w} (s) = \gamma (bs)^{\gamma} e^{bs} \Gamma (-\gamma,bs)
= 1 - (bs)^{\gamma} e^{bs} \Gamma (1-\gamma,bs)\:,
\end{equation}
with $b:=\gamma/a$ and the incomplete Gamma function $\Gamma(a,x) :=
\int_{x}^{\infty} dt \; e^{-t} t^{a-1}$.

For $\gamma=1$ the Laplace transform of the waiting time distribution is
\begin{equation}
\label{Laplace_times_PDF_g1}
\tilde{w} (s) = 1 - \frac{2s}{a} e^{\frac{2s}{a}} E_{1}(\frac{2s}{a}) \quad ,
\end{equation}
where $E_1(x)$ is the exponential integral,
$E_1(x):=-ei(-x)=\int_{x}^{\infty} dt \frac{e^{-t}}{t}$.

Here we are only interested in the long time behavior of the MSD,
which corresponds to taking the limit $s \rightarrow 0$ in
Eqs.~(\ref{Laplace_times_PDF}), (\ref{Laplace_times_PDF_g1}) with
$\gamma$ being constant. For $0 < \gamma < 1$, taking the lowest order
of the expansion we get $\Gamma (1 - \gamma,bs) \simeq \Gamma (1 -
\gamma)$ and
\begin{equation}
\label{wait}
\tilde{w}(s) \simeq 1 - (bs)^{\gamma} \Gamma(1-\gamma) \quad .
\end{equation}
Eq.~(\ref{Laplace_MSD2}) then yields
\begin{equation}
\tilde{ \left< x^2 \right> } = \frac{1}{\Gamma(1-\gamma)b^{\gamma}} s^{-1-\gamma}
\end{equation}
and its inverse Laplace transform is
\begin{equation}
\label{MSD}
\left< x^2 \right> = \frac{a^{\gamma} \sin (\pi \gamma)}{2 \pi \gamma^{1+ \gamma}} t^{\gamma},
\; \; \; 0 < \gamma < 1\quad .
\end{equation}
Similarly, for the normal diffusive case $\gamma > 1$ we get from Eq.\ 
(\ref{Laplace_times_PDF}) as the small $s$-asymptotics of the waiting
time distribution
\begin{equation}
\label{wait_normal}
\tilde{w}(s) \simeq 1 - \frac{bs}{\gamma-1}
\end{equation}
Using Eq.~(\ref{Laplace_MSD2}) and the inverse Laplace transform
yields
\begin{equation}
\label{MSD_norm}
\left< x^2 \right> = \frac{\gamma-1}{b} \; t, \; \; \; \gamma > 1\quad .
\end{equation}
In the remaining case $\gamma=1$, the expansion of the exponential
integral is $ei(-x)=\ln (x) + \tilde{\gamma}$, where $\tilde{\gamma}$
is Euler's constant. For small $s$ Eq.~(\ref{Laplace_times_PDF_g1})
reads
\begin{equation}
\label{wait_g1}
\tilde{w}(s) \simeq 1 + bs \ln (bs)
\end{equation}
and Eq.~(\ref{Laplace_MSD2}) gives
\begin{equation}
\label{Laplace_MSD_g1}
\tilde{ \left< x^2 \right> } = \frac{b^{-1} s^{-2}}{\ln (1/bs)}-\frac{1}{s}\quad .
\end{equation}
The inverse Laplace transform of the first term can be obtained by
applying Karamata's Tauberian theorem \cite{Bingham}, which relates
the Laplace transform of a function given in the form $f(s) \sim
A\frac{\Gamma(1+\gamma)}{s^{\gamma+1}} L(1/s)$ for $s \rightarrow 0$
and $\gamma > -1$ to the inverse $F(t) \sim A t^{\gamma}L(t)$ for $t
\rightarrow \infty$. The function $L$ must meet the requirement of
being slowly varying, $\frac{L(ux)}{L(x)} \rightarrow 1$ for $u>0$ as
$x \rightarrow \infty$. In our case we have $\gamma=1$, $\Gamma(2)=1$
and the function $L(1/s)=\frac{1}{\ln (1/bs)}$ satisfies the above
condition. For the MSD we thus get
\begin{equation}
\label{x2g1}
<x^2> = \frac{a t}{\ln(a t)}\quad , \quad \gamma = 1\quad .
\end{equation}

In summary, we obtain for the GDC 
\begin{equation}
\label{CTRW_Geisel}
K = \begin{cases}
\frac{a^{\gamma} \sin (\pi \gamma)}{\pi \gamma^{1+\gamma}}\quad , & 0 < \gamma < 1 \cr
0\quad , & \gamma = 1 \cr
a\frac{\gamma - 1}{\gamma} \quad , & \gamma > 1
\end{cases}
\end{equation}
with $\gamma=\frac{1}{z-1}$. For $\gamma = 1$ we get $K=0$ because of
the logarithmic term in Eq.~(\ref{x2g1}).

Now we compare this result with our numerical simulations. As an
example, $K$ as a function of $a$ determined by
Eq.~(\ref{CTRW_Geisel}) is shown in Fig.\ \ref{Z3} (a), where $z=3$.
For $z>2$, the difference to the result of Ref.\ \cite{Gei84} is by a
factor of two. This yields approximately the right slope for small
$a$, however, it does not reproduce the trivial value of $K(4)=0$ when
particles cannot escape from the unit interval anymore. Hence, further
modifications are necessary.

\section{Modified CTRW theory}\label{mod_theory}

The easiest way to modify standard CTRW theory is by varying either
the waiting time or the jump PDF. However, the waiting time
distribution is straightforwardly determined by the model as a power
law for all intermittent values of $z$.  This suggests to change the
jump PDF Eq.~(\ref{jump_PDF}) only, which in a first attempt one may
write as \begin{equation}
  \label{real_jump_PDF1} \lambda (x) = \frac{p}{2} \delta (\left| x
  \right| - 1) + (1 - p) \delta (x) \quad .
\end{equation} 

\begin{figure}[t] \centering
  \includegraphics[width=0.48\textwidth]{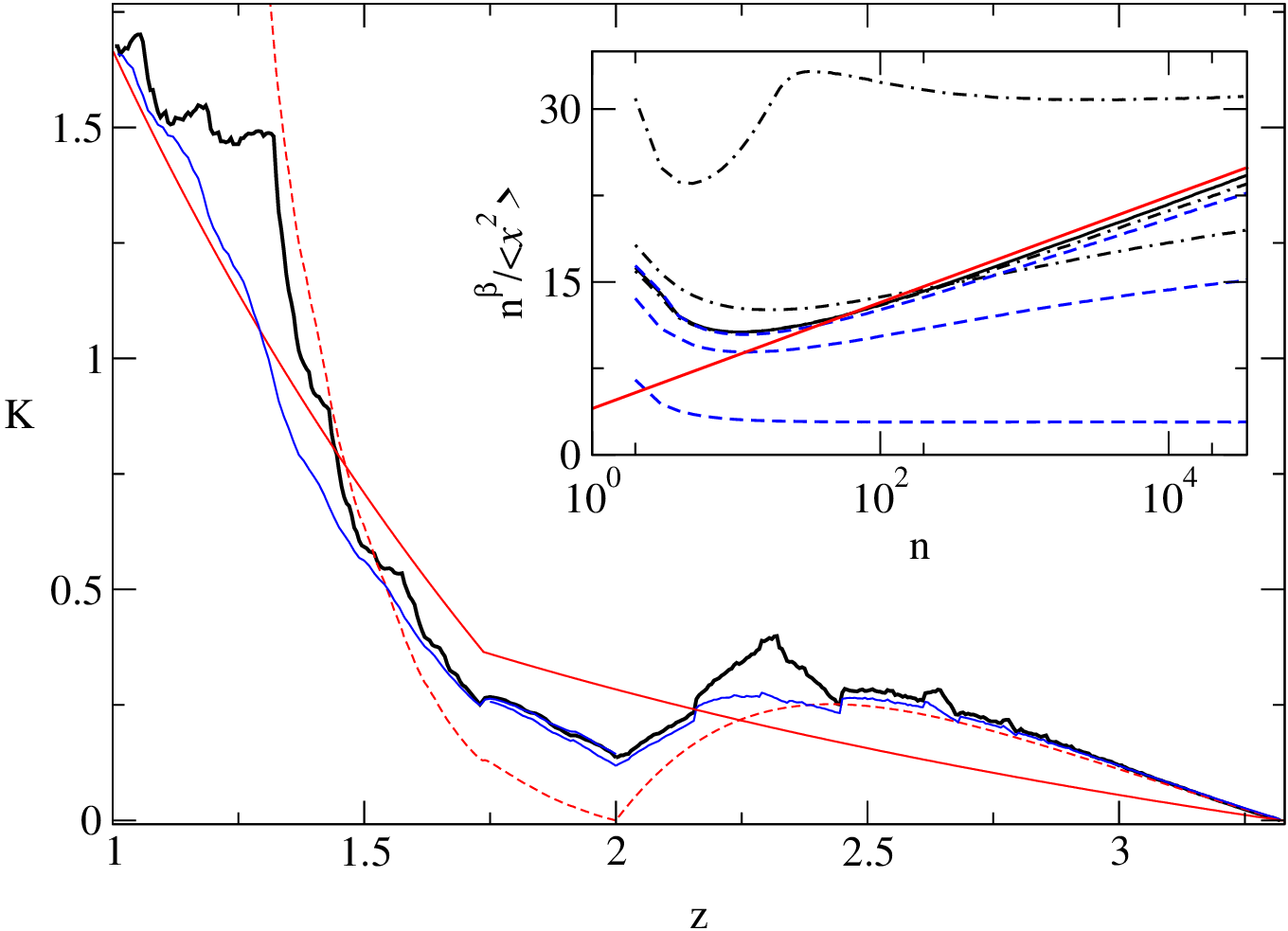}\\ 
  \caption{$K$ as a function of $z$ for $a = 5$. The bold black 
    line depicts again the computer simulation results from
    Fig.~\ref{K_z}.  The thin line that is on top at $z=2$ represents
    the simple random walk approximation Eq.~(\ref{eq:dknorm}) for
    $i=2$.  The dashed line corresponds to the modified CTRW
    approximation Eqs.~(\ref{int_jump_length}), (\ref{mod_GDC_CTRW}).
    The thin line that is in the middle at $z=2$ displays the first
    term of the TGK formula Eq.~(\ref{GDC_GK}).  The inset reveals the
    existence of logarithmic corrections in the MSD by plotting
    $n^{\beta}/\left< x^2 \right>$ with respect to $\ln n$ for
    $a=3.14$ and different $z$ close to the transition point at $z=2$.
    The dashed lines correspond to $z=1.5, 1.9, 1.95$ from bottom to
    top, the dashed-dotted lines to $z=2.5, 2.1, 2.05$ from top to
    bottom. The thick solid line represents $z=2$, the thin solid line
    is proportional to $\ln n$ as a guide for the eye.}
  \label{K_z_ctrw}
\end{figure}

Here the second term reflects the fact that the particle can stay on
the unit interval with probability $(1 - p)$. By assuming that the
density of particles is uniform on the unit interval, $p$ is
determined by the size of the escape region, $p(a) = 2 \Delta(a)$ with
$\Delta = \frac{1}{2} - x_c$, where $x_c$ is the solution of the
equation $x_c + a x^{z}_{c} = 1$. For normal diffusion it is
well-known that $K_{rw} = p(a)$ provides a random walk approximation
for the diffusion coefficient which is asymptotically correct in case
of nearest neighbor jumps, $p(a)\ll 1$
\cite{Kla95,Kla96,KlDo97,KlKo02}. For farther than nearest neighbor
jumps this approximation is straightforwardly generalized by weighting
the integer jump distance squared with the respective probability,
based on the escape region, to perform such a jump
\cite{Kla02}. Assuming that one knows about the power law time
dependence of the MSD Eq.~(\ref{GDC_Def}) for this map, one can apply
the same approximation to anomalous diffusion.  In Fig.\
\ref{K_z_ctrw} the result is shown for $K(z)$ at $a=5$. As one can
see, this simple argument indeed roughly reproduces the coarse
functional form of $K(z)$ and yields the exact value at $z=1$.

Accordingly, the modification of standard CTRW theory
Eq.~(\ref{real_jump_PDF1}) works for nearest neighbor jumps only.  We
thus further amend the jump pdf Eq.~(\ref{real_jump_PDF1}) by
introducing a typical jump length $l$, \begin{equation}
\label{real_jump_PDF} \lambda (x) = \frac{p}{2} \delta (\left| x
\right| - l) + (1 - p) \delta (x) \quad , \end{equation} with Fourier
transform $\hat{\lambda} (k) = p \cos(lk) + 1 - p$.  This jump length
we define either by the actual mean displacement 
\begin{equation}
\label{jump_length} l_1 = \left\{ | M(x) - x | \right\} 
\end{equation}
or, under the simplifying assumption of integer jumps, by the
coarse-grained displacement 
\begin{equation} \label{int_jump_length}
l_{2} = \left\{ | [M(x)] | \right\} \quad , 
\end{equation} 
where the square brackets hold for the largest integer less than the
argument. Here the averages, denoted by curly brackets, are performed
starting from a uniform initial ensemble of walkers with respect to
the conditional probability density that particles leave the unit
interval. The latter ensemble averages are then averaged over time.
Repeating the calculations from the previous section with
Eq.~(\ref{real_jump_PDF}), we obtain the MSD
\begin{equation} 
\label{mod_x2} \left< x^2 \right> = p l_{i}^2
\begin{cases} \frac{a^{\gamma} \sin (\pi \gamma)}{\pi
\gamma^{1+\gamma}} \; t^{\gamma}\quad , & 0 < \gamma < 1 \cr
\frac{at}{\ln (at)}\quad , & \gamma = 1 \cr \frac{\gamma-1}{\gamma} \;
at\quad , & \gamma > 1 \quad , \end{cases} 
\end{equation} $i=1,2$,
where the typical jump length $l_i$ is given by
Eqs.~(\ref{jump_length}), (\ref{int_jump_length}). Hence, our final
CTRW expression for the GDC reads \cite{fnote3}
\begin{equation} 
\label{mod_GDC_CTRW} K =
p l_{i}^2 \begin{cases} \frac{a^{\gamma} \sin (\pi \gamma)}{\pi
\gamma^{1+\gamma}} \quad , & 0 < \gamma < 1 \cr 0\quad , & \gamma = 1
\cr a\frac{\gamma-1}{\gamma}\quad , & \gamma > 1 \quad .  \end{cases}
\end{equation} 
Fig.\ \ref{Z3} (a) shows the CTRW approximation
Eq.~(\ref{mod_GDC_CTRW}) with jump length $l_1$,
Eq.~(\ref{jump_length}), which describes the coarse dependence of the
GDC very well over a large range of parameters.  The CTRW result with
integer jump length $l_2$, Eq.~(\ref{int_jump_length}) with
Eq.~(\ref{mod_GDC_CTRW}), is shown in Fig.\ \ref{Z3} (b). It gives an
asymptotically exact approximation of the GDC for small values of $a$,
however, it also holds well for larger parameters.  A
quantitative comparison moreover indicates (within numerical accuracy)
that this combination yields results which exactly correspond to our
numerical findings for $K$ at all integer values of
$h=M(\frac{1}{2})>0$, which for $z=3$ are equivalent to
$a=4,12,20,...$. This generalizes results obtained previously for the
normal diffusion coefficient at $z=1$ \cite{Kla96,KlDo97}.

We now focus on $K$ as a function of $z$. Fig.~\ref{K_z_ctrw} shows
the modified CTRW approximation Eqs.~(\ref{int_jump_length}),
(\ref{mod_GDC_CTRW}) in comparison with simulation results. In the
strongly anomalous regime of large $z$ the approximation reproduces
the GDC from simulations very well, however, for smaller values of $z$
there are obvious deviations. Let us first explain the problem for
$z\to 1$, the region around $z=2$ will be discussed in the following
section.

Let us recall that according to Eqs.~(\ref{GDC_Def}),(\ref{alfa}), for
$z<2$ the map exhibits normal diffusion in the long-time limit. One
can further split this regime into $3/2\le z<2$, where the MSD shows
transient anomalous diffusion which becomes normal for long times
while for $1\le z<3/2$ it represents purely normal dynamics
\cite{Gas88}. However, in the latter case it is well-known that the
waiting time distribution is exponential \cite{Kla96},
\begin{equation}
w(t)=\tilde{\gamma} e^{-\tilde{\gamma} t}\quad , \label{eq:wtexp}
\end{equation}
where $\tilde{\gamma}$ is the escape rate, and not a power law such as
the CTRW approximation Eq.~(\ref{times_PDF2}). In fact, in the limit of
$z\to1$ Eq.~(\ref{mod_GDC_CTRW}) leads to $K=apl_i^2$, whereas the
correct random walk result for normal diffusion reads
\cite{Kla96,KlDo97}
\begin{equation} K=pl_i^2 \quad . \label{eq:dknorm} 
\end{equation} 
Note that for $z\to1$ and by assuming a uniform density on the unit
interval, in case of $l_2\ll2$ this equation yields $K=p$, whereas for
$l_2\gg2$ we get $K=l_2^2$ thus recovering the two simple random walk
results mentioned earlier. Eq.~(\ref{eq:dknorm}) is indeed
straightforwardly obtained from CTRW theory if one repeats the
calculations with the exponential waiting time distribution
Eq.~(\ref{eq:wtexp}) and $\tilde{\gamma}=1$, i.e., by assuming that on
average a particle leaves the unit cell after one time-discrete
jump. We thus conclude that the CTRW result Eq.~(\ref{mod_GDC_CTRW})
gives only a reliable approximation for the GDC if $\gamma\le 2$,
whereas for $\gamma>2$ the simple random walk result for normal
diffusion Eq.~(\ref{eq:dknorm}) should be used. This is in line with
Fig.~\ref{K_z_ctrw}.

\section{Suppression of the GDC at a dynamical transition}\label{sup}

Let us now understand the behavior of $K(z)$ around $z=2$, where the
map exhibits a dynamical transition \cite{Beck,Wang} from normal
to anomalous diffusion, see
Eqs.~(\ref{GDC_Def}),(\ref{alfa}). Interestingly, this transition is
clearly visible in the strength of the GDC, which to our knowledge has
not been reported before. We now explain this phenomenon in detail.

As mentioned in Section \ref{NumRes}, around the transition point
there are significant deviations between CTRW theory and the
simulation results. These differences are displayed in
Fig.~\ref{K_z_ctrw}.  At $z=2$ the CTRW approximation forms a
non-differentiable little wedge with $K(2)=0$ at the minimum, whereas
the simulations yield $K(2)>0$. We recall that by increasing the
computation time there is some slow convergence of the simulation data
towards the CTRW solution, see the inset of Fig.~\ref{K_z}, but that
we were not able to achieve quantitative agreement with the CTRW
prediction of $K(2)=0$.

This very slow convergence of the simulation results as well as the
logarithmic term in the MSD at $z=2$, see Eq.~(\ref{mod_x2}), suggest
that the deviations between simulations and CTRW theory are due to
logarithmic corrections in the MSD for parameter values around $z=2$.
Curiously, such terms are not present in Eq.~(\ref{mod_x2}) for $z\neq
2$ ($\gamma\neq1$).  By refining our CTRW analysis, we will now show
that such logarithmic terms indeed exist. However, for $z\neq 2$ they
hold for large but finite times only, whereas for $z=2$ they persist
in the limit of infinite time. As is obvious from Eqs.~(\ref{mod_x2}),
(\ref{mod_GDC_CTRW}), for $z=2$ the surviving logarithmic term leads
to a full suppression of the GDC, whereas close to the transition
point finite-time logarithmic corrections yield a gradual suppression
of the strength of diffusion.

We start again from the exact expression for the Laplace transform of
the waiting time PDF $\tilde{w} (s)$,
Eq.~(\ref{Laplace_times_PDF}). Combining this equation with the jump
PDF Eq.\ (\ref{real_jump_PDF}) and Eqs.\ (\ref{Montroll_Weiss}),
(\ref{Laplace_MSD}) we obtain the Laplace transform of the MSD
\begin{equation} \label{sup_Laplace_MSD2} 
\tilde{\left< x^2 \right>} = pl_i^2 b^{-\gamma} s^{-1-\gamma}e^{bs}
\Gamma^{-1}(1-\gamma,bs) - \frac{pl_i^2}{s}\quad .
\end{equation} 
We immediately drop the second term, since its inverse Laplace
transform is simply constant. Let us focus now on the limits $s\to0$ and
$\gamma\to 1$. In case of $0<\gamma\le1$ the $\Gamma$ function in
Eq.~(\ref{sup_Laplace_MSD2}) can be expanded to
\begin{equation}
\label{s1}
\Gamma (1-\gamma,bs) = \Gamma (1-\gamma) - \frac{(bs)^{1-\gamma}}{1-\gamma} -
\sum_{j=1}^{\infty} \frac{(-1)^{j}(bs)^{j+1-\gamma}}{(j+1-\gamma)j!}\quad .
\end{equation}
From this equation it follows that as $\gamma$ approaches $1$ from
below one may consider not only the first term as in
Eq.~(\ref{wait}) but also the second one, since both terms are
divergent for $\gamma\to 1^-$. All other terms are nonsingular 
and can be safely neglected. We further expand the second
term of Eq.~(\ref{s1}) to
\begin{eqnarray}
\frac{(bs)^{1-\gamma}}{1-\gamma} &=& \frac{e^{(1-\gamma)\ln (bs)}}{1-\gamma}\nonumber \\
&=& \frac{1}{1-\gamma} + \ln (bs) + \sum_{j=2}^{\infty}
\frac{(1-\gamma)^{j-1}}{j!} \; (\ln (bs))^j\quad . \nonumber \\ 
\label{s2} 
\end{eqnarray}
Note that here we have obtained a series of logarithmic terms which,
as we shall see, provides the logarithmic corrections we are looking
for.  By imposing the condition that $|(1-\gamma)\ln (bs)| < 1$, we
may keep only the first two terms of Eq.~(\ref{s2}),
\begin{equation}
\label{s4}
\frac{(bs)^{1-\gamma}}{1-\gamma} \simeq \frac{1}{1-\gamma} + \ln (bs) \quad .
\end{equation}
Finally, we expand the first term of Eq.~(\ref{s1}) to
\begin{equation}
\label{s3}
\Gamma (1-\gamma) = \frac{1}{1-\gamma} - \tilde{\gamma} +
\sum_{j=2}^{\infty} \frac{(1-\gamma)^{j}}{j!} \int_{0}^{\infty} dt \; e^{-t} (\ln t)^j \quad ,
\end{equation}
where $\tilde{\gamma}$ is again Euler's constant.  Because we are
interested in the small $\gamma$ limit, we keep only the first term of
this expansion. In summary, we obtain for the Laplace transform of the
MSD Eq.~(\ref{sup_Laplace_MSD2}) in the limits of $|(1-\gamma)\ln (bs)| < 1$
and $\gamma\to 1^-$
\begin{equation} 
\tilde{\left< x^2 \right>} = pl_i^2 b^{-\gamma} s^{-1-\gamma} \;
\frac{1}{\ln (\frac{1}{bs})}\quad .
\end{equation}
Inversion of the Laplace transform yields as the final result
\begin{equation}
\label{sup_MSD_log}
\left< x^2 \right> =
\frac{pl_i^2 t^{\gamma}}{b^{\gamma} \Gamma(1+\gamma) \ln (t/b)}, \quad  t < t_{cr} 
\end{equation}
with $t_{cr}:= b e^{1/(1-\gamma)}$. For $t \gg t_{cr}$ and $0<\gamma<1$
one can drop the second term in the $\Gamma$ function expansion Eq.\
(\ref{s1}), and the asymptotic CTRW result Eq.~(\ref{mod_x2}) is
recovered. Interestingly, $t_{cr}$ diverges when $\gamma \to 1^-$, and
at $\gamma=1$ one thus arrives at the asymptotic $t/\ln t$ dependence of
Eq.~(\ref{mod_x2}).

Analogous corrections are obtained for $1 < \gamma$ near the dynamical 
transition point, $\gamma \to 1^+$. By applying the same arguments
as above we get
\begin{equation}
\label{sup_MSD_log3}
\left< x^2 \right> = \frac{pl_i^2 t}{b^{\gamma} \Gamma(1+\gamma) \ln (t/b)},\quad t < \tilde{t}_{cr}\quad ,
\end{equation}
where $\tilde{t}_{cr}:= b e^{1/(\gamma -1)}$.  In the long time limit
we again recover the CTRW asymptotics Eq.~(\ref{mod_x2}) \cite{fnote}.

These analytical findings are in agreement with results from computer
simulations. In the inset of Fig.~\ref{K_z_ctrw}, $n^{\beta}/\left<
  x^2 \right>$ is shown as a function of $\ln n$ close to the
transition point at $z=2$ for different values of $z$ and $a=3.14$.
The logarithmic corrections are less obvious if $z$ is sufficiently
different from $2$, where $n^{\beta}/\left< x^2 \right>$ quickly
converges to $1/K$. However, the logarithmic corrections are getting
significant when $z$ approaches the dynamical transition value. We have
thus identified the precise dynamical origin of the suppression of the
GDC at $z=2$.

Dynamical transitions are quite ubiquitous in intermittent maps
and have been widely discussed in the literature in terms of the time
dependence of the MSD. However, it appears that so far no attention
has been paid to a possible critical behavior of the associated
GDC. Our results lead us to conjecture that suppression and
enhancement of the GDC are typical for dynamical transitions in
anomalous dynamics altogether \cite{Zum93a}.

\section{Time-fractional equation for subdiffusion}\label{FFP}

The great success of the CTRW approach is related to the fact that it
not only predicts the power $\gamma $ but also the form of the coarse
grained PDF $P(x,t)$ of displacements \cite{Zum93}. Correspondingly
the anomalous diffusion process generated by our model is not
described by an ordinary diffusion equation but by a fractional
generalization of it. Starting from the Montroll-Weiss equation and
making use of the expressions for the jump and waiting time PDFs
Eqs.~(\ref{times_PDF2}), (\ref{jump_PDF}), we rewrite
Eq.~(\ref{Montroll_Weiss}) in the long-time and -space asymptotic form
\begin{equation} 
s^{\gamma}\hat{\tilde{P}} - s^{\gamma - 1} = - \frac{p l_i^2}{2 c b^{\gamma}} k^2
\hat{\tilde{P}} \label{Montroll-Weiss_2} 
\end{equation} 
with $c = \Gamma (1-\gamma)$ and $b=\gamma /a$.  For the initial
condition $P(x,0)=\delta (x)$ of the PDF we have $\hat{P}(k,0)=1$. It
is now helpful to recall the definition of the Caputo fractional
derivative of a function $G$ \begin{equation} \frac{\partial^{\gamma}
G}{\partial t^{\gamma}} := \frac{1}{\Gamma (1-\gamma)} \; \int_0^t
dt^{^{\prime }}(t-t^{^{\prime }})^{-\gamma }\frac{\partial G}{\partial
t^{^{\prime }}} \label{Caputo} \end{equation} and its Laplace
transform \cite{Podl,Mai97},
\begin{equation}
\int_{0}^{\infty} dt \; e^{-st} \frac{\partial^{\gamma} G}{\partial
t^{\gamma}} = s^{\gamma} \tilde{G} (s) - s^{\gamma - 1} G(0) \quad .
\label{Caputo_Laplace} 
\end{equation}
Noticing that the left part of Eq.~(\ref{Montroll-Weiss_2}) precisely
coincides with the Laplace transform of the Caputo derivative of the
PDF and turning back to real space, we arrive at the time-fractional
diffusion equation
\begin{equation}
\frac{\partial ^\gamma P(x,t)}{\partial t^\gamma } = D \; \frac{\partial ^2P}{\partial x^2}
\label{Frac_Dif_Eq}
\end{equation}
with $D$ given by the modified CTRW theory Eqs.~(\ref{K}),
(\ref{mod_GDC_CTRW}).  Time-fractional diffusion equations of such a
form have already been extensively studied by mathematicians
\cite{Gor00}.  Note that in application to our model another version
of such an equation was proposed in Ref.~\cite{Bar03}, which uses a
Riemann-Liouville fractional derivative. It can be easily shown that
both forms of time-fractional diffusion equations are equivalent under
rather weak assumptions \cite{ACunp}.  Yet two other forms of
subdiffusive fractional equations (however, not with applications to
maps) were proposed in Refs.~\cite{Sai97} and \cite{Met00}.  Again,
after some recasting these equations yield Eq.~(\ref{Frac_Dif_Eq})
\cite{LUnp}.

The solution of Eq.\ (\ref{Frac_Dif_Eq}) is expressed in terms of an
M-function of Wright type \cite{Mai97},
\begin{equation}
P(x,t)=\frac 1{2\sqrt{D}t^{\gamma /2}} M \left( \xi,\frac \gamma 2\right) 
\label{Sol}
\end{equation}
with $\xi :=|x|/\sqrt{D}t^{\gamma/2}$, where we use the representation
of the M-function
\begin{equation}
M (z, \gamma) = \frac{1}{\pi} \sum_{i=1}^{\infty} \frac{(-z)^{i-1}}{(i-1)!} 
\Gamma (\gamma i) \sin (\gamma i \pi)\quad .
\label{M_func}
\end{equation}
This solution gives exactly the same asymptotics that was obtained in
Ref.~\cite{Zum93} for small and large values of $\xi$,
\begin{equation}
P(x,t) \simeq t^{-\gamma/2} \begin{cases}
1 - a_1 \xi + a_2 \xi^2 \quad , & \xi \ll 1 \cr
\xi^{\frac{\gamma - 1}{2-\gamma}} \exp(-b_1 \xi^{\frac{2}{2-\gamma}}) \quad , & \xi \gg 1 \quad ,
\end{cases}
\label{Sol2}
\end{equation}
\begin{figure}[t] \centering
\includegraphics[width=0.48\textwidth]{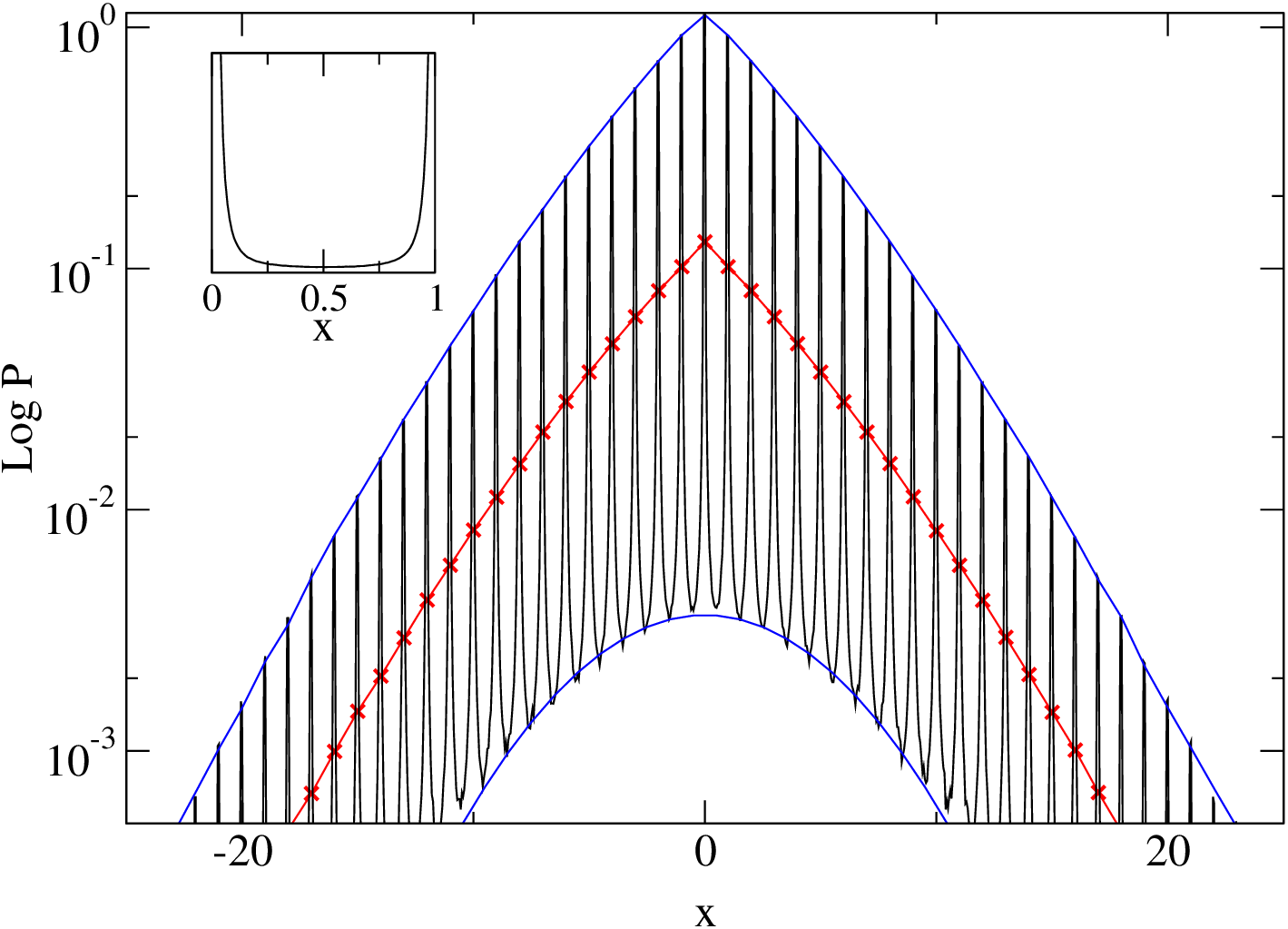}\\
\caption{Main figure: The oscillatory structure shows the PDF 
    obtained from simulations of the map Eq.~(\ref{map_eq}) for $z =
    3$ and $a = 8$.  The simulated PDF was computed from an ensemble
    of $10^7$ trajectories after $n = 10^3$ iterations. The crosses
    (x) represent the simulated PDF coarse-grained over unit
    intervals.  The continuous line in the middle yields the
    analytical solution of the fractional diffusion equation
    Eq.~(\ref{Sol}) with the same $z$ while $D$ is taken from
    simulations.  The upper and the lower curves correspond to fits
    with an M-function form, calculated with the same $z$ but for
    different $D$, and a Gaussian form, respectively. More
    explanations are provided in the text, see Sec.\ref{FFP} below.
    The inset depicts schematically the simulated PDF for the map
    Eq.~(\ref{map_eq}) modulo $1$.}
  \label{PDF}
\end{figure}
where $a_1, a_2, b_1$ are some constants that are given explicitly in
Ref.\ \cite{Zum93}.

Similar to the modified CTRW approximation of the parameter-dependent
GDC we now find that the analytical PDF Eq.~(\ref{Sol})
describes only the coarse scale of the PDF obtained from simulations
but not its fine structure.  This mismatch can be understood
  related to the fact that our model consists of a periodic lattice
defined by specific elementary cells. Any elementary cell
develops a characteristic ``microscopic'' PDF, characterized by
the map Eq.~(\ref{map_eq}) restricted onto the unit interval, which
exhibits singularities at the marginally stable fixed points. The
inset of Fig.~\ref{PDF} schematically depicts such a PDF of an
elementary cell after a large number of iterations.  Precisely this
functional form yields the building block for the PDF of the
periodically continued map, see the main part of Fig.~\ref{PDF}. If
one eliminates the fine structure by averaging over whole unit
intervals, one obtains a coarse-grained PDF that is in excellent
  agreement with the analytical solution of the fractional diffusion
  equation Eq.~(\ref{Frac_Dif_Eq}), see Fig.~\ref{PDF}.

Such an interplay between the ``microscopic'' PDF of a single
scatterer and the ``macroscopic'' PDF of the spatially extended system
was already reported for the normal diffusive model at $z=1$
\cite{Kla96,Kla02,Kla04} and was recently also studied in billiards
\cite{Sand05}. However, in case of anomalous diffusion the PDF
exhibits a further remarkable property: Fig.\ \ref{PDF} shows that the
oscillatory structure of the whole PDF is bounded by two different
functions. The upper curve is of M-function type and connects
all local maxima of the microscopic PDF. These maxima are
situated in regions of the map where the dynamics is regular, 
due to the marginally stable fixed points. The lower curve, on the
other hand, is Gaussian and is determined by all local minima of
the microscopic PDF. These minima are generated in regions of the map
being far away from the marginally stable fixed points, where
the dynamics is locally most strongly chaotic \cite{fnote2}.
Fig.\ \ref{PDF} thus nicely exemplifies the microscopic origin of
anomalous diffusion in terms of intermittency.

\section{Taylor-Green-Kubo approach and fractal functions}\label{ff}

We now turn back to the parameter dependence of the GDC. In Section
\ref{mod_theory} we have shown that an amended CTRW theory correctly
describes the coarse dependence of the model's GDC, whereas the
fractal fine structure is not captured by this approach. This reflects
the fact that CTRW theory is a purely {\em stochastic} approach
involving a randomness assumption between jumps, see the
approximations involved as outlined in Section \ref{theory} and
\ref{mod_theory}, whereas the origin of the fractality of the GDC lies
in the existence of long-range {\em deterministic} dynamical
correlations, see the analysis in Section \ref{NumRes}.

This motivates us to propose an alternative approach for analyzing the
GDC, which is based on the Taylor-Green-Kubo (TGK) formula
\cite{Tay21,Kubo} for diffusion in maps. This theory has successfully
been applied to the fractal diffusion coefficient of the normal
diffusive map at $z=1$, where it could partly be worked out
analytically \cite{Kla96}, to a nonlinear map with a complicated
bifurcation scenario \cite{Kor02} and to normal diffusion in billiards
\cite{KlKo02}. An advantage of this analysis is that it systematically
incorporates dynamical correlations. A disadvantage is that, in
contrast to CTRW theory, for the map under consideration we could
implement it only numerically.

The basic idea is to generalize the TGK formula for maps to anomalous
diffusion. Following the usual derivation \cite{Tay21,KlKo02}, the
first step is to transform the MSD in Eq.~(\ref{GDC_Def}) into sums
over increments, or velocities, $v_k:=x_{k+1}-x_k$,
\begin{equation}
\label{MSD2}
<(x_n - x_0)^2> =
<\sum_{k=0}^{n-1} v_k \sum_{l=0}^{n-1} v_{l}>\: (n\to\infty)\:.
\end{equation}
For normal diffusion the second step requires to define the ensemble
average expressed by the angular brackets in terms of the invariant
PDF obtained for the map restricted onto the unit interval.  In case
of the map Eq.~(\ref{map_eq}) this still works for $z<2$, where the
dynamics of the spatially extended map is normal diffusive.  However,
for $z\ge2$ it is well-known that the map of the elementary cell does
not possess a non-trivial invariant PDF anymore.  Mathematical
analysis \cite{Hu95} shows that in this case there exist two
physically relevant invariant measures, one which is concentrated on
the marginally stable fixed points and one that lies in-between on the
interval $(0,1)$. The first measure has still nice, so-called SRB
properties and yields a PDF in form of a $\delta$-distribution on the
marginally stable fixed points.  The second one, however, is not
normalizable anymore and hence is called an infinite invariant
measure. In other words, if one starts a computer simulation from an
ensemble of points uniformly distributed on the unit interval, the
underlying stochastic process is not stationary and the PDF of an
elementary cell, computed by a histogram method, does not converge in
time to a well-defined invariant PDF \cite{Zum93a,Gas88}.

Consequently, in contrast to normal diffusion Eq.~(\ref{MSD2}) cannot
further be simplified by using time-translational invariance, and for
the GDC Eq.~(\ref{K}) we have to stop at
\begin{equation}
K = \lim_{n
\rightarrow \infty} \frac{1}{n^{\beta}} \left[ <\sum_{k=0}^{n-1} v_k^2> + 2
<\sum_{k=0}^{n-1} \sum_{l=1}^{n-1} v_k v_{k+l}> \right] \:.
\label{GDC_GK}
\end{equation}
In numerical simulations we find that the first term alone is already
proportional to $n^{\beta}$. If the system is ergodic, as for $z<2$
in our model, this term boils down to $n <v_0^2>$, and by neglecting
any higher-order terms in Eq.~(\ref{GDC_GK}) we recover the random
walk result for normal diffusion \cite{Kla96,KlKo02,KlDo97}. For $z\ge
2$ the situation is again more complicated. Here only generalized
ergodic theorems may hold \cite{Lutz04,Hu95}, which is intimately
related to the existence of infinite invariant measures as outlined
above.  Whether in this case a CTRW result such as
Eq.~(\ref{mod_GDC_CTRW}) can be extracted from the first term is a
non-trivial open question.

\begin{figure}[t]
\centering
\includegraphics[width=0.48\textwidth]{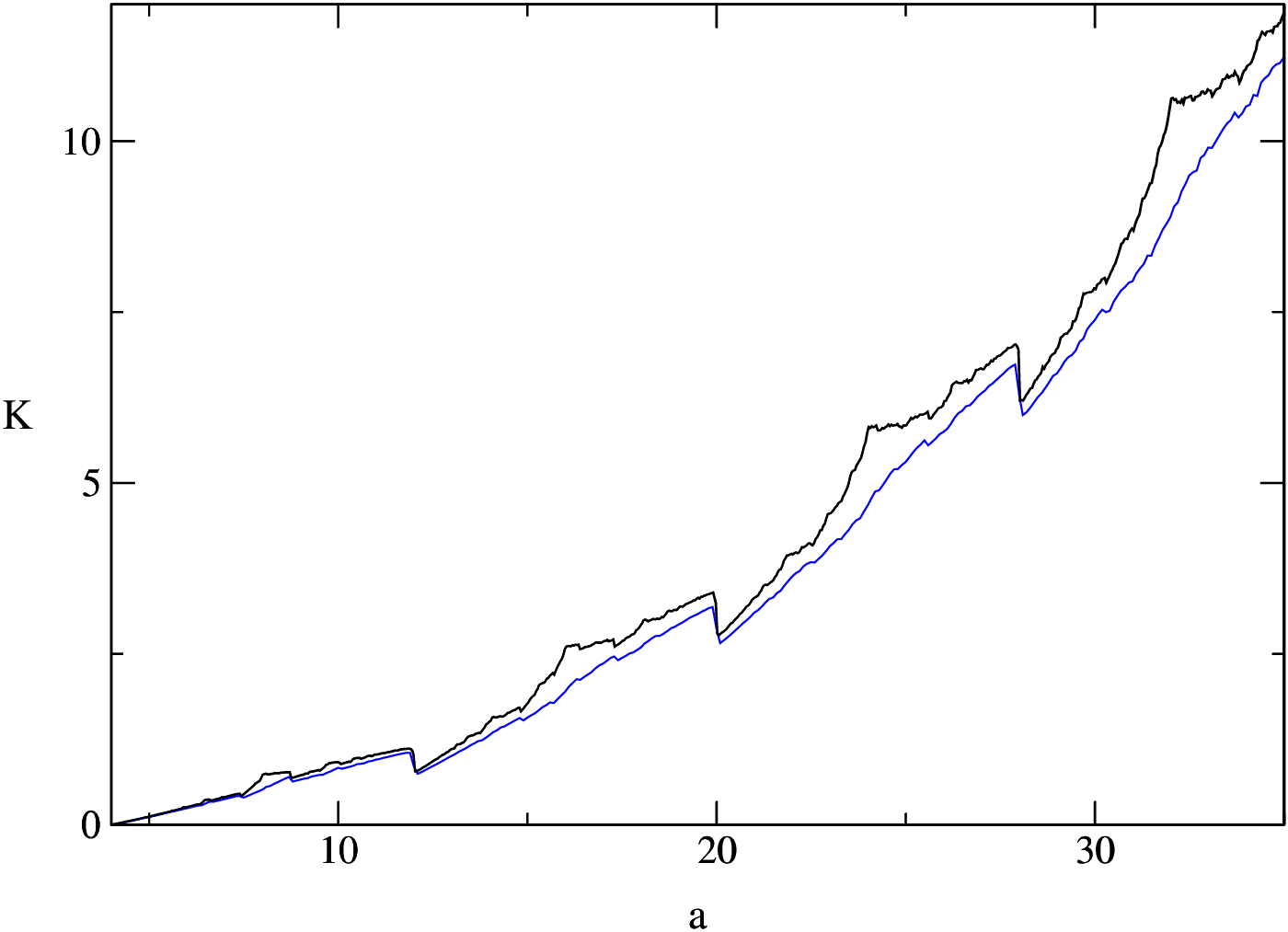}\\
\caption{Comparison of the first term of the  
  anomalous TGK formula Eq.~(\ref{GDC_GK}) (lower line) with the
  computer simulation results from Fig.~\ref{Z3} (a).} \label{GK}
\end{figure}

However, considering the first term only as an approximation of $K$
for all $z$, the numerical results are depicted in Fig.~\ref{GK}. The
comparison of this approximation with the previous simulation results
shows that this term provides a first little step beyond a modified
CTRW approach, since it reproduces the major irregularities of $K$ as
a function of $a$ and even follows the suspected discontinuities
discussed in Section \ref{NumRes}. However, our numerical precision is
not sufficient to conclude whether it yields exact values for $K$ at
integer heights.

Eq.~(\ref{GDC_GK}) thus provides a suitable starting point for a
systematic evaluation of the fractal structure of the GDC. Since the
series expansion in Eq.~(\ref{GDC_GK}) is exact, working out further
terms one should recover more and more structure in the GDC on fine
scales \cite{KlKo02,Kor02}.  Interestingly, the second term in
Eq.~(\ref{GDC_GK}) can be understood as a velocity autocorrelation
function that, in contrast normal diffusion, depends on two time
scales. The existence of a second time scale points to the phenomenon
of {\em aging} in dynamical systems \cite{Bar03,Rad04}. In our case
this can be understood as a consequence of infinite invariant
measures.

\begin{figure}[t]
\centering
\includegraphics[width=0.48\textwidth]{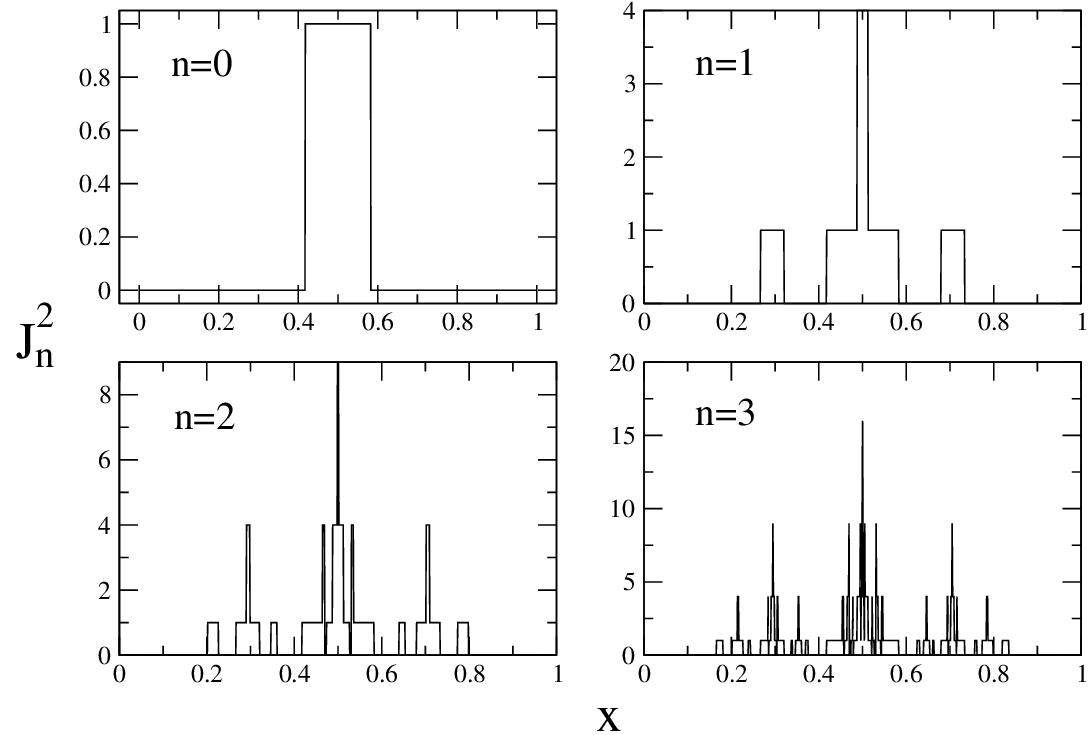}\\
\caption{The first four iterations of the jump function $J^2_n(x)$ defined by 
  Eq.~(\ref{J}) for $z=3$, $a=8$, $n=0, 1, 2, 3$ based on a uniform
  ensemble of $N=10^3$ initial conditions. Note that there emerges a
  complicated fine structure.}
\label{Jump_Func}
\end{figure}

\begin{figure}[t]
\centering
\includegraphics[width=0.48\textwidth]{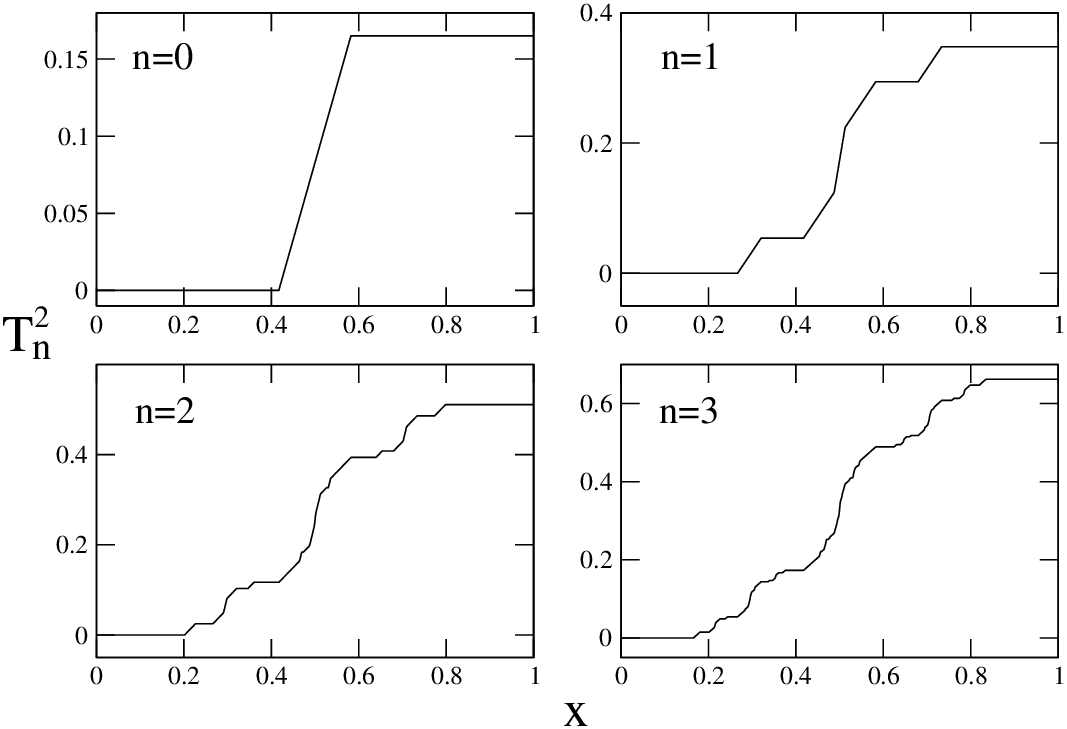}\\
\caption{The first four iterations of the generalized de Rham-type function 
  $T_n^2(x)$ as defined by Eq.~(\ref{T}), which is obtained by
  numerically integrating $J_n^2(x)$ depicted in Fig.~\ref{Jump_Func}.
  The parameter values are the same as in Fig.~\ref{Jump_Func}. Note
  that there emerges a fractal structure.}
\label{Anom_Takag}
\end{figure}

Here we do not further pursue these questions but focus instead onto a
direct link between the GDC and fractal functions as provided by the
TGK formula, see Refs.\ \cite{Kla96, KlKo02,Kor02} for normal
diffusion.  As was shown in Ref.~\cite{KlKo02}, Eq.~(\ref{GDC_GK})
also holds if the velocities $v_k$ are replaced by the integer
velocities $j_k := [x_{k+1}] - [x_k]$. It is now useful to start from
$K$ expressed again by Eq.~(\ref{MSD2}), trivially rewritten as
\begin{equation}
\label{GDC_GK2}
K = \lim_{n \to \infty} \frac{1}{n^{\beta}} <\sum_{k=0}^{n-1} j_k \sum_{l=0}^{n-1} j_{l}>\quad .
\end{equation}
The two sums suggest to define the jump function
\begin{equation}
J_n(x) := \sum_{k=0}^n j_k\quad ,
\label{J}
\end{equation}
which satisfies the recursion relation \cite{Kla96}
\begin{equation}
J_n(x) = j_0(x) + J_{n-1}(M(x)) \quad .
\label{J_recursion}
\end{equation}
The jump function gives the integer value of the displacement of a
particle after $n$ time steps, which started at some initial position
$x\equiv x_0$. It thus contains essential information about the
microscopic scattering process of a particle by showing how
sensitively the displacement depends on initial conditions.
Eqs.~(\ref{GDC_GK2}), (\ref{J}) imply
\begin{equation}
K = \lim_{n \to \infty} \frac{1}{n^{\beta}} <J_{n-1}^2(x)>\quad .
\end{equation}
Fig.~\ref{Jump_Func} displays the product of jump functions
$J_n^2(x)$, which governs the anomalous diffusion process, for
a representative parameter value. Clearly, as time evolves the
structure of this function is getting more complicated. The GDC, in
turn, is determined by the cumulative function
\begin{equation}
T_n^2(x) := \int_{0}^{x} dy \; J^2_{n}(y) \quad ,
\label{T}
\end{equation}
where we fix the integration constants by the condition that $T_n^2(0)
= 0$ and by requiring that the whole function is continuous.
Integration of Eq.~(\ref{J_recursion}) would yield a recursive
functional equation for $T_n^2(x)$, which is of the same type as the
one derived in Ref.~\cite{Kor02}. The solutions of such equations are
generalized de Rham-, respectively generalized Takagi functions
\cite{Kla96,Kor02}. However, at present such generalized de
Rham-equations cannot be solved analytically, hence we do not work out
the details but instead compute $T_n^2(x)$ directly from simulations,
according to the definition Eqs.~(\ref{J}), (\ref{T}).  Results for
the first four iterations of $T_n^2(x)$ are shown in
Fig.~\ref{Anom_Takag}.  With respect to the construction of
$T_n^2(x)$, it may not come too much as a surprise that in the limit
of $n\to\infty$ this function exhibits a fractal structure.
Interestingly, the GDC is obtained by integrating over this structure.
If one starts from a uniform ensemble of initial conditions the result
reads
\begin{equation}
K = \lim_{n \to \infty} \frac{1}{n^{\beta}} T_n^2(1)\quad .
\label{KT}
\end{equation}
This equation relates the GDC to a fractal function representing the
microscopic scattering process of our model. The numerical result for
the GDC Eq.~(\ref{KT}) is in good agreement with the one employing the
MSD. In the light of Fig.~\ref{Anom_Takag} and Eq.~(\ref{KT}), it may
not be too surprising anymore that under parameter variation a highly
non-trivial GDC is obtained from this model.

\section{Conclusions}\label{con}

In this paper we have studied subdiffusion generated by a paradigmatic
one-dimensional intermittent map. In contrast to previous works here
we focused onto the parameter dependence of the GDC, which is an
anomalously diffusive generalization of the diffusion coefficient
known for normal diffusion.  Computer simulations suggested that this
GDC is a fractal function of control parameters.  This finding was
corroborated by a qualitative explanation of the fractal structure in
terms of complicated sequences of forward- and backward scattering
(turnstile dynamics), which are topologically unstable under parameter
variation. Our analysis furthermore led us to conjecture that the GDC
is a discontinuous function of control parameters.

In trying to understand the coarse functional form of the GDC, we
applied standard CTRW theory to our model. By suitably amending
previous calculations, we arrived at analytical approximations that
enabled us to reproduce the whole coarse functional form of the GDC
yielding asymptotically exact results in the limit of large and small
parameter values. However, there are clear deviations between this
theory and simulations around a dynamical transition from normal
to anomalous diffusion.  By refining our amended theory, we were able
to explain these deviations in terms of logarithmic corrections
leading to ultra-slow convergence of our simulation results and
eventually yielding a full suppression of the GDC right at the
transition point. These findings were confirmed by simulations of the
MSD.

We then studied in detail the PDF of our model. We first derived a
time-fractional subdiffusion equation from CTRW theory.  The
coarse-grained PDF of our model turned out to be in excellent
agreement with the non-Gaussian solution of our fractional diffusion
equation. On a fine scale, however, the simulation results yielded an
oscillatory structure reflecting the microscopic details of the
intermittent scattering process. This structure was generated by the
invariant density of the single scatterers hence revealing an
interesting interplay between microscopic scattering and macroscopic
diffusion.

A more detailed understanding of the GDC was finally provided in terms
of an anomalous TGK formula, which again is a generalization of the
TGK formula for normal diffusion.  The structure of this formula
points at intimate relations to infinite invariant measures, aging,
and generalized de Rham-type fractal functions. All these terms define
very active topics of research. We thus hope that, along these lines,
in future work more detailed relations between mathematical infinite
ergodic theory and the physics of anomalous diffusive processes can be
established.

In the present paper we have only studied a one-dimensional
subdiffusive map, however, we expect these findings to be typical for
spatially extended, low-dimensional, anomalous deterministic dynamical
systems altogether. Further studies may also focus on the relation
between the anomalous TGK formula and CTRW theory, on a spectral
analysis of the Frobenius-Perron operator of this model, and on
analyzing a superdiffusive map by similar methods.

The authors gratefully acknowledge helpful discussions with J.
Dr\"ager, R.  Gorenflo, J.  Klafter, R. Metzler and A. Pikovsky. They
thank the MPIPKS Dresden for hospitality and financial support. AVC
acknowledges financial support from the DFG.



\begin{thebibliography}{10}
\bibitem{fnote3} Note that in the corresponding formula of
Ref.~\cite{Unp}, $\gamma$ should have been defined correctly by
Eq.~(\ref{eq:defgamma}) of the present paper.
\bibitem{fnote2} The upper curve in Fig.\ \ref{PDF} was obtained from
  $P(x)=a_0 \exp (- x^2/a_1)$, the lower one from $b_0 \;
  \tilde{M}(\left| x \right|/ b_1, \frac{\gamma}{2})$. The fit
  parameters were found to be $a_0=0.0036, a_1=55.0183$ and $b_0 =
  3.05, b_1 = 0.37$.
\end{thebibliography}
\end{document}